\documentclass[a4paper,fleqn,usenatbib]{mnras}
\usepackage{CJK}
\usepackage{graphicx}        
\usepackage{natbib}         
\usepackage{amssymb}        
\usepackage{amsmath}
\usepackage{color}           
\usepackage{url}             
\usepackage[normalem]{ulem}
\renewcommand{\vec}[1]{ {\mathbf #1} }

\newcommand{\dd}{\mathrm{d}}

\title[Current helicity and anisotropy]{Current helicity and magnetic field anisotropy in solar active regions}
\author[H. Xu et al.]{H.~Xu${^1}$\thanks{E-mail: xhq@bao.ac.cn}, R.~Stepanov${^2}$, K.Kuzanyan$^{1,3}$, D.~Sokoloff$^{1,4}$,  H.~Zhang$^{1}$ and
Y.~Gao$^{1}$\\
$^{1}$Key Laboratory of Solar Activity, National Astronomical
Observatories, Chinese Academy of Sciences, Beijing 100012, China\\
$^{2}$Institute of Continuous Media Mechanics, Korolyov str. 1,
614061 Perm, Russia\\
 $^{3}$ IZMIRAN, Troitsk, Moscow 142190, Russia\\
$^{4}$ Department of Physics, Moscow University, 119992 Moscow,
Russia }

\begin{document}


\date{Accepted XXX. Received YYY; in original form ZZZ}

\maketitle

\begin{abstract}
The electric current helicity density {\bf $\displaystyle
\chi=\langle\epsilon_{ijk}b_i\frac{\partial b_k}{\partial
x_j}\rangle$ }contains six terms, where $b_i$ are components of the
magnetic field. Due to the observational limitations, only four of
the above six terms can be inferred from solar photospheric vector
magnetograms. By comparing the results for simulation we
distinguished the statistical difference of above six terms for
isotropic and anisotropic cases. We estimated the relative degree of
anisotropy for three typical active regions and found that it is of
order 0.8 which means the assumption of local isotropy for the
observable current helicity density terms is generally not satisfied
for solar active regions. Upon studies of the statistical properties
of the anisotropy of magnetic field of solar active regions with
latitudes and with evolution in the solar cycle, we conclude that
the consistency of that assumption of local homogeneity and isotropy
requires further analysis in the light of our findings.

\end{abstract}


\section{Introduction}
     \label{S-Introduction}

Solar magnetic cycle is believed to be excited by solar dynamo
mechanism based on a joint action of differential rotation and
mirror-asymmetric convection. Differential rotation known from
helioseismology produces toroidal large-scale magnetic field from
poloidal one while mirror asymmetric convection is responsible for
transformation of the {\bf toroidal} large-scale magnetic field into
{\bf poloidal} one in order to close the chain of self-excitation.
Simple symmetry arguments show that a link between the toroidal and
poloidal magnetic fields must be governed by a mirror-asymmetric
{\bf quantity (generally a pseudo tensor)} , however, there are
several ways on how to implement this link in particulars (e.g. by
notion of cyclonic motions, \citet{Park55}; or by magnetic tubes
arising and being twisted by Coriolis force, as in
\citet{Babcock1961}, \citet{Leighton69} mechanism, what gives a
variety of solar dynamo models. From the other hand, the degree of
mirror asymmetry is believed to be moderate and the very degree of
mirror asymmetry seems to be hardly determined from observational
data. The point is that in order to quantify the mirror asymmetry of
a dynamo one has to know 3D distribution of the mean-field
characteristics and compute their spatial derivatives.

For example, a more straightforward quantity known as hydrodynamic
(or kinetic) helicity density $ \langle {\bf v} \cdot {\rm curl\,}
{\bf v}  \rangle    $ which determines the excess of right-hand
helixes against left-hand ones requires 3D distribution of velocity
field $\bf v$ and its derivatives while conventional Doppler-effect
gives line-of-sight velocities only. Note that we are interested in
averaging quantities and $ \langle \dots    \rangle $ means
corresponding averaging. While in mean-field dynamo theory the
averaging is carried out over the ensembles of turbulent pulsations,
practically given the observational data in vector magnetograms of
solar active regions, we average by area in the available
field-of-view of something similar to that.

A practically accessible way to quantify mirror asymmetry
observationally was firstly suggested by \cite{Seehafer90}. He
pointed out that vector magnetographic observations provide three
magnetic field components on a surface $z={\rm const}$ at solar
atmosphere (local coordinates $x$ and $y$ are parallel to the solar
surface, and for a limited field of view we can ignore the effect of
curvature). He used the force-free field parameter $\alpha$ as a
proxy of electric current helicity. The available vector magnetic
field is not sufficient to calculate the entire current helicity
density
 $\chi=  \langle {\bf B} \cdot {\rm curl\,} {\bf B}    \rangle    $
 and
to quantify relative number of right handed twisted magnetic tubes against
left-handed ones, however, one can calculate a mirror-asymmetric
quantity

\begin{equation}
\chi_{z} =  \langle B_z ({\rm curl\,} {\bf B})_z    \rangle     =
\left \langle B_z \left( {{\triangledown_x B_y} } -
{{\triangledown_y B_x} } \right) \right \rangle  \,  , \label{See}
\end{equation}
which makes one of the three believed to be similar contributions to
the entire quantity $\chi$, i.e., $\chi_{x}$ and $\chi_{y}$ (they
can be obtained from Eq.~(\ref{See}) by circular replacement of
indices). Note that $\chi_{z}$ do not contain derivative in
direction $z$. If magnetic field $\bf B$ is locally statistically
isotropic, all three contribution in $\chi$ are statistically equal
and the natural consequence is
\begin{equation}
\chi= 3\chi_{z} \,. \label{one_third}
\end{equation}

The force-free field parameter $\alpha$ and the current helicity
parameter $\chi$ become available for observational determination in
solar active regions \citep[see, e.g.,][]{pev94, Abramenko96, Bao98,
Hagino04}. The above statistical studies show that the sign of
$\alpha$ and $\chi$ is same, which is predominantly negative in the
northern hemisphere and positive in the southern hemisphere. Results
of monitoring $\chi_{z}$ in two last solar cycles and butterfly
diagrams for solar cyclic variation of this quantity are presented
in \citet{Zhang10}. The quantity followed the helicity polarity rule
as well as pronounced areas on the butterfly diagrams where the
polarity rule is inverted \citep{Zhang10}. The result looks
instructive for solar dynamo modeling \citep{Zhang12}. Dynamo
interpretation of the observational data is usually based on
assumption of the local statistical isotropy, and so $\chi_{z}$ is
considered as an observational tracer for $\chi$.

Conventional theory of turbulence is originated from \cite{Ko41},
while \cite{Iro63} and \cite{Krai65} ideas presumes that the
velocity and magnetic fields become statistically homogeneous in
sufficiently small scales (though \citet{Goldreich97} have stressed
the role of magnetic field anisotropy in MHD turbulence). It is
however not clear what scales in the solar photospheres can be
considered as homogeneous enough for this assumption and whether
typical active regions may fall under this consideration. A
perspective to verify to what extent the solar magnetic field at the
scale of active regions can be considered as statistically isotropic
has not been yet considered observationally at least in the approach
suggested by \citet{Seehafer90}: the point is that calculation of
other contributions to the current helicity to be compared with
$\chi_{z}$ would require derivatives of magnetic field components in
$z$ direction while in fact we have three components of the magnetic
field at surface $z={\rm const}$ only.

The aim of this paper is to show that one can refurnish the approach
of \citet{Seehafer90} in a way to verify the hypothesis of local
isotropy in the scales of active regions. Instead of to present
$\chi$ as a sum of three contributions $\chi_{z}$, $\chi_{x}$ and
$\chi_{y}$ which have to be equal in a locally isotropic case, we
present this quantity as a sum of six quantities to be equal in the
isotropic case. Four out of the six quantities do not contain
derivatives in $z$ direction and are potentially accessible for
observations.

We demonstrate that the available bulk of data for the magnetic
field vector in solar active regions enable us to obtain
statistically robust estimates for the four quantities. Local
statistical isotropy implies two pairs of identities for the above
four quantities. We show that one pair of the identities holds while
the other fails being confronted with the observational data. We may
believe that the magnetic field occurs to be substantially
statistically anisotropic in scales of active regions. As a result,
the quantity $\chi_{z}$ which has been traced for last two solar
cycles has to be considered as a specific mirror asymmetric tracer
of anisotropic solar MHD rather than purely $\chi/3$. We discuss the
importance of this conclusion for solar dynamo models.

\section{Observational Data}
   \label{Data}

We used 6629 vector magnetograms observed by Solar Magnetic Field
Telescope (SMFT) at Huairou Solar Observing Station from 1988 to
2005.  This  data sample was used earlier by \citet{Gao08} and
\citet{Zhang10}. The SMFT is equipped with a birefringent filter for
wavelength selection and KD*P crystals to modulate polarization
signals. The Fe 1 5324.19 \AA\ line is used. A vector magnetogram is
built using four narrow-band (0.125\AA) filtergrams of Stokes
\textit{I, Q, U} and \textit{V} parameters. The center wavelength of
the filter can be shifted and is normally at -0.075 \AA\ for the
measurements of longitudinal magnetic field and at the line center
for the transversal magnetic fields \citep{Ai86}. The $180^{\circ}$
ambiguity in the azimuth angle ($\phi$) was resolved following
\citet{Wang94} by comparison with a potential field. We used the
method given by \citet{Gao08} to correct the Faraday rotation to the
azimuthal angles for vector magnetic field.

We have also established the levels of noise for the longitudinal
and transverse components of the magnetic field as 20 G and 100 G,
in accord with previous works on analysis of Huairou SMFT data
\citep[e.g.,][]{Abramenko96}. That values have been used to estimate
the impact of noise in statistical studies as it has been performed
earlier by e.g. \cite{Bao98} or \citet{Zhang10}.

\section{The Method}
     \label{S-Method}

\begin{figure}
{ \centering\includegraphics[width=0.5\textwidth]{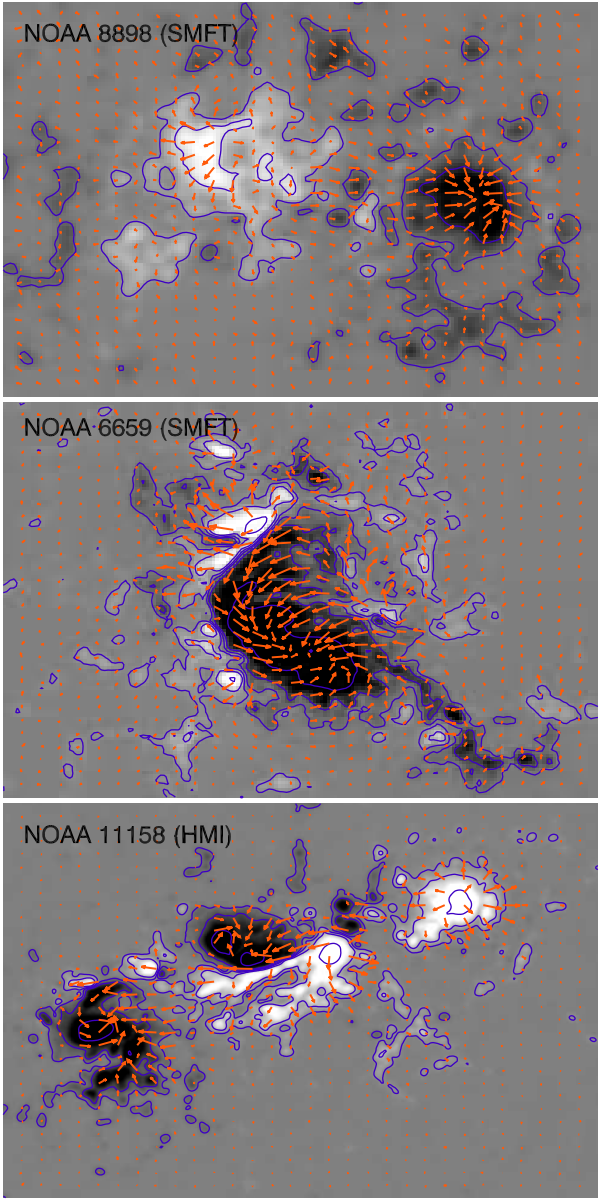}}
\caption{%
The vector magnetograms for three examples: NOAA 8898 (S13.0W7.0)
observed by SMFT on 01:34 UT, March 8, 2000 (top); NOAA 6659 (N28.6W4.5)
observed by SMFT on 05:29 UT, June 9, 1991 (middle); NOAA 11158 (S19.0E11.0)
observed by HMI on 23:59 UT, Feb 14, 2011 (bottom). The arrows represent the
direction of transverse field and the contours represent the
longitudinal magnetic fields of $\pm 100, 400, 1600, 3200 G$.}
\label{fig4}
\end{figure}

By definition the current helicity is a scalar product of the
magnetic field pseudo-vector and its curl (proportional to the
electric current) vector. Generally speaking this construction can
be obtained as a trace of a more generic three-index tensor quantity
as it has been considered by \citet{Zhang12}, see their Appendix 1.
For our analysis, let us denote local observable quantities by low
cases, i.e. the local magnetic field ${\bf b}$ which is observable
at the solar surface $z=0$ as well as the local electric current
helicity density $h_{\rm c}= {\bf b} \cdot {\rm curl\,} {{\bf b}}$.
We denote $(x,y)$ local Cartesian coordinates on the image plane.
The direction to the observer $z$ on the plane is fixed $z=0$. We
reserve above mentioned notations $\bf B$ and $\chi$ for the
magnetic field and current helicity in homogeneous and isotropic
model. Then according to definition of ${\rm curl\,}$ this quantity
naturally comprises of six parts: $\displaystyle
\chi=\epsilon_{ijk}b_i\frac{\partial b_k}{\partial
x_j}=h_1+h_2+h_3+h_4+h_5+h_6$ where

\begin{eqnarray}
h_1 = b_{z}\left(\frac{\partial b_y}{\partial x}\right);  &\quad& h_2 = b_{z}\left(-\frac{\partial b_x}{\partial y}\right);
 \nonumber
 \\
h_3 = b_{x}\left(\frac{\partial b_z}{\partial y}\right); &\quad&  h_4 =b_{x}\left(-\frac{\partial b_y}{\partial z}\right);
\\
h_5 = b_{y}\left(\frac{\partial b_x}{\partial z}\right); &\quad& h_6 =b_{y}\left(-\frac{\partial b_z}{\partial x}\right)
\,.
\nonumber
\end{eqnarray}
We denote integral quantities over the available magnetogram
field-of-view by capital cases $H_i=\int {h_i} dxdy$ for $i=1\,{\rm
to}\,6$, so the overall average current helicity reads
\begin{eqnarray}
H_{\rm c} &=&
H_1+H_2+H_3+H_4+H_5+H_6 \\
&=& \int b_{z}\left(\frac{\partial b_y}{\partial x}\right) dx dy
+\int b_{z}\left(-\frac{\partial b_x}{\partial y}\right) dx dy
\nonumber
\\
&+& \int b_{x} \left(\frac{\partial b_z}{\partial y}\right) dx dy
+ \int b_{x}\left(-\frac{\partial b_y}{\partial z}\right) dx dy
\nonumber
\\
&+& \int b_{y}\left(\frac{\partial b_x}{\partial z}\right) dx dy
+ \int b_{y}\left(-\frac{\partial b_z}{\partial x}\right) dx dy
\,.
\nonumber
\label{H123456}
\end{eqnarray}

\begin{table*}
\begin{center}
\caption{Summary of illustration computations of the helicity and
boundary integrals in formula (\ref{int_by_parts}). The unit of
$H_{1}$, $H_{6}$ and boundary integral is $10^{14} G^{2}m$. }
\label{tab:xu1}
\begin{tabular}{ccccccccccc}
\hline
Instr. &NOAA & date & Time & coordinate & $H_{1}$ & $H_{6}$ &  $\%$ difference & bound.int.  & err.est. \\
\hline
SMFT &8898 & 2000.03.08& 01:37 &S13.0W7.0 & $-0.0493$ & $-0.0495$  & 2.87\%& 0.0014  & 0.0135\\
SMFT &6659 & 1991.06.09 & 05:29 & N28.6E4.5 & $-1.4978$ & $-1.4983$ & 0.31\% & $-0.0046$ & 0.0306 \\
HMI  & 11158 & 2011.02.14 & 23:47 & S20W17 & 0.0793 & 0.0784 & 3.48\% & $-0.0027$  & 0.0123\\
\hline
\end{tabular}
\end{center}
\end{table*}

Using the integral by parts formula, we can obtain the following
relation between helicity parts, for example,
\begin{eqnarray}
\int b{z}\left(\frac{\partial b_y}{\partial x}\right) dx dy = \int
b_{y}\left(-\frac{\partial b_z}{\partial x}\right) dx dy +
\int\limits_\Gamma b_{z}b_{y} d{\it {l}} \label{int_by_parts}
\end{eqnarray}
where the latter integral is taken over contour $\Gamma$ at the
boundary of our field of view (boundary integral). If we assume the
magnetic field at the boundary of the active regions is weak enough,
that means the boundary integral is very small, then the averages of
the corresponding parts helicity are approximately equal in pairs
$H_1 \simeq H_6$ and $H_2 \simeq H_3$. We illustrate validity of
this consideration below.

For the purpose of illustration of our data we selected two active
regions observed by SMFT in 1999 and 2000, and one to compare with
other instruments, we used the data for an active region observed by
Helioseismic and Magnetic Imager onboard the Solar Dynamics
Observatory (HMI/SDO) in 2011. Figure~\ref{fig4} shows the vector
magnetograms for three active regions observed by SMFT and HMI/SDO.
We can see that the magnetic field is nearly potential for NOAA 8898
and the magnetic field is strong helical for NOAA 6659. The spatial
resolution of HMI is higher than SMFT. We calculated the integrals
$H_{1}$ and $H_{6}$ and the boundary integrals in formula
({\ref{int_by_parts}}). The results are listed in Table 1.  The
boundary integral is about 2.87\%, 0.31\% and 3.48\% of the mean
value of $H_{1}$ and $H_{6}$ for active region NOAA 8898, 6659 and
11158 respectively. In order to estimate tolerable error in
computation of the boundary integral we use the typical noise
levels: 20 G in $B_{z}$ and 100 G in $B_{y}$.  For example, active
region NOAA 6659, the error is about 2.04\% of the mean value of
$H_{1}$ and $H_{6}$. The corresponding boundary integral is well
below the error level. These observational examples show that the
above assumption of validity of integration by parts works pretty
well when the magnetic field at the boundaries of the field of view
is weak, so the formula (\ref{int_by_parts}) of integration by parts
results in these equalities. Therefore, we established on the basis
of both theoretical consideration  and observational illustration
that the four parts of observable current helicity are equal by
pairs to their counterparts: $H_1 \approx H_6$ and $H_2 \approx
H_3$.

\begin{figure} 
\includegraphics[width=0.45\textwidth]{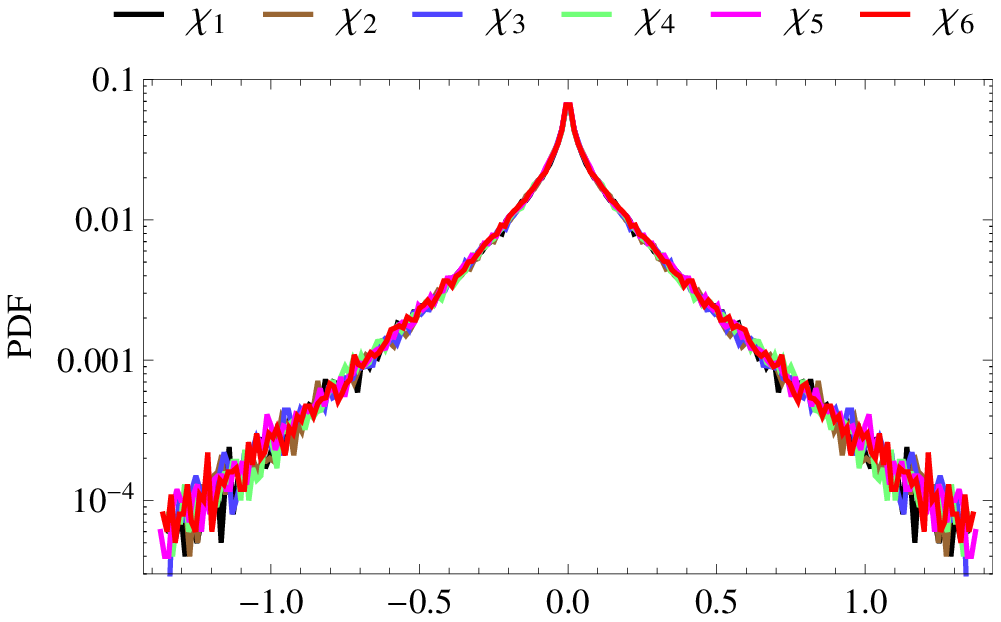}
\includegraphics[width=0.45\textwidth]{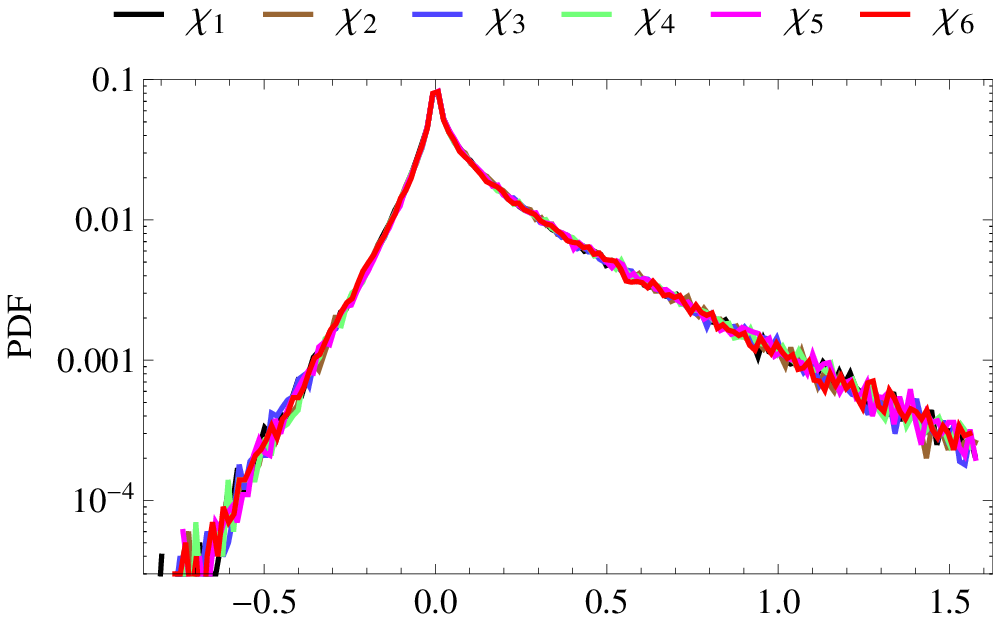}
\caption{%
PDF for the six helicity parts: non-helical isotropic case (top),
helical isotropic case (bottom). Note that only four out of the six
are of observational interest. The x-axis is the relative value of
helicity parts and the y-axis is the probability density function.
All parts statistically coincide with each other though for the
helical case there is a pronounceable bias visible as asymmetry over
the center of distribution.} \label{fig1}
\end{figure}

It is trivial to see that the local values of all helicity parts are generally unequal.
Now let us consider whether the integral identities implied to mean helicity parts hold for the observational data.

First of all, all the parts of the total current helicity mentioned
in the Introduction can be expressed in notations analogous to the
ones in formulae (\ref{H123456}) and their observational
counterparts
\begin{eqnarray}
\chi_z\,\,=\,\, \chi_1+\chi_2 \,\,\,  \longleftrightarrow \,\,\,
H_1+H_2=H_{{\rm c}z} \, , \nonumber
\\
\chi_x\,\,=\,\, \chi_3+\chi_4 \,\,\,  \longleftrightarrow  \,\,\,
H_3+H_4=H_{{\rm c}x} \, ,
\\
\chi_y\,\,=\,\, \chi_5+\chi_6 \,\,\,  \longleftrightarrow  \,\,\,
H_5+H_6=H_{{\rm c}y} \,, \nonumber
\end{eqnarray}
\noindent and notice that only the former part can be fully computed
from observations as it does not contain derivatives with respect to
$z$. Only one term in each of the two latter parts can be computed
and the two terms $H_4$ and $H_5$ are not available from
observations on the image plane. We either cannot use formula
(\ref{int_by_parts}) for them as the relevant derivative is with
respect to $z$ but integration is carried out over $x$
and $y$.

Let us assume local statistical isotropy of turbulence and take for instance two parts of helicity equal, say

\begin{equation}
\chi_{z}=\chi_{x} \, \nonumber
\end{equation}

Then we immediately have that as $H_1+H_2=H_3+H_4$, and due to formula (\ref{int_by_parts}) $H_1=H_6$ and $H_2=H_3$, then the three parts are equal

\begin{equation}
H_1=H_4=H_6 \,.
\label{h146}
\end{equation}

\noindent
Therefore, an additional assumption on the equality of the other helicity parts

\begin{equation}
\chi_{x}=\chi_{y} \nonumber
\end{equation}

would automatically lead to equality of the other three parts

\begin{equation}
H_2=H_3=H_5\,.
\label{h235}
\end{equation}

The above consideration means that for verification of the assumption of local isotropy unobservable parts of helicity $H_4$ and $H_5$ need to be evaluated in order to check equations (\ref{h146}-\ref{h235}).

Before doing further analysis we are going to see what relationship
between the parts of helicity we can expect from theoretical
consideration. For that purpose model simulation of the helicity
parts is performed in the next section.

\section{Simulations}

\begin{figure*} 
{
\includegraphics[width=0.48\textwidth]{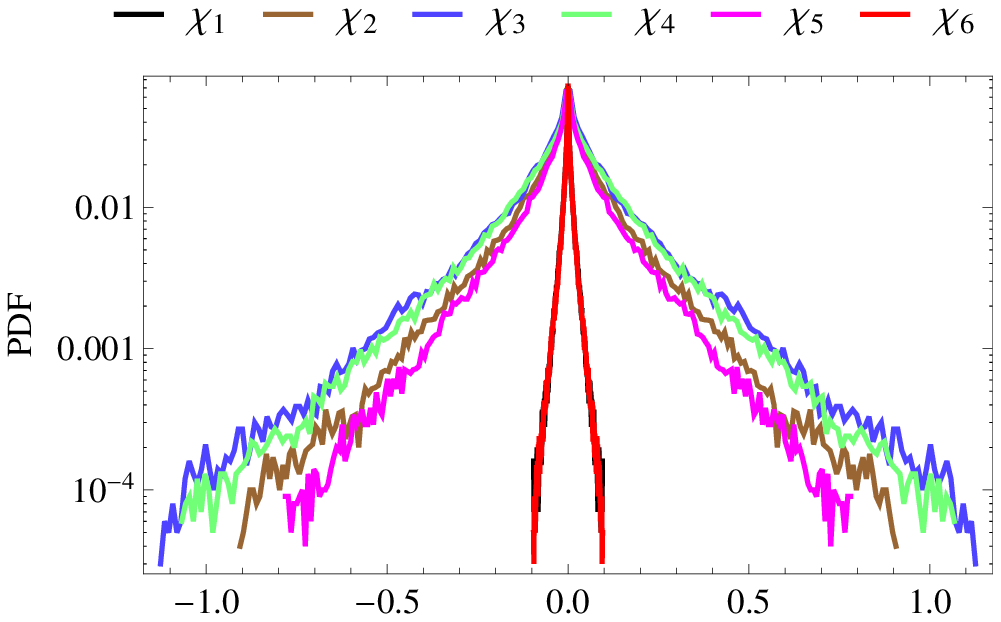}
\hspace{4mm}
\includegraphics[width=0.48\textwidth]{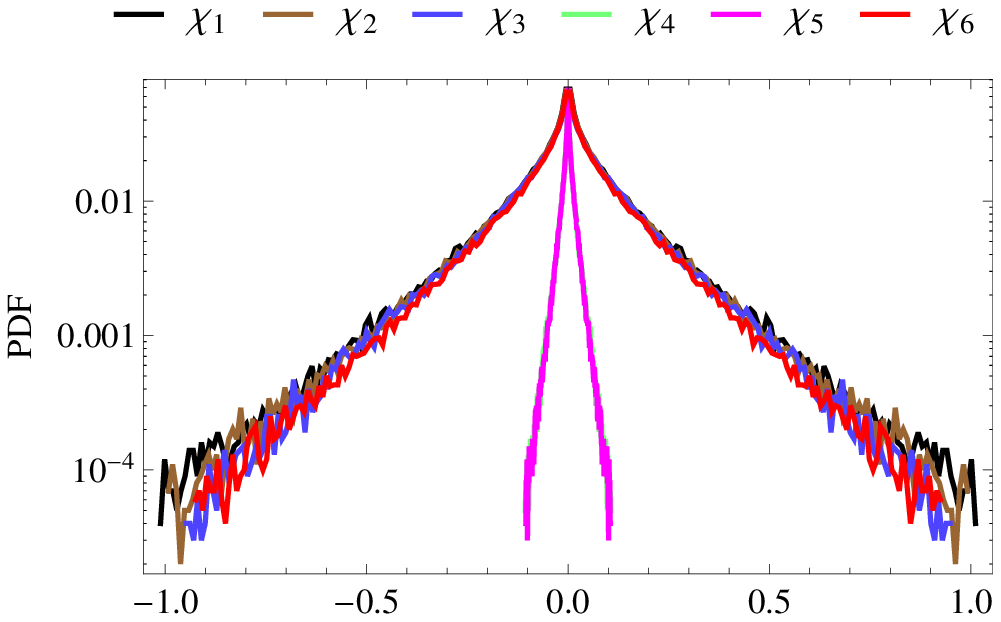}
\includegraphics[width=0.48\textwidth]{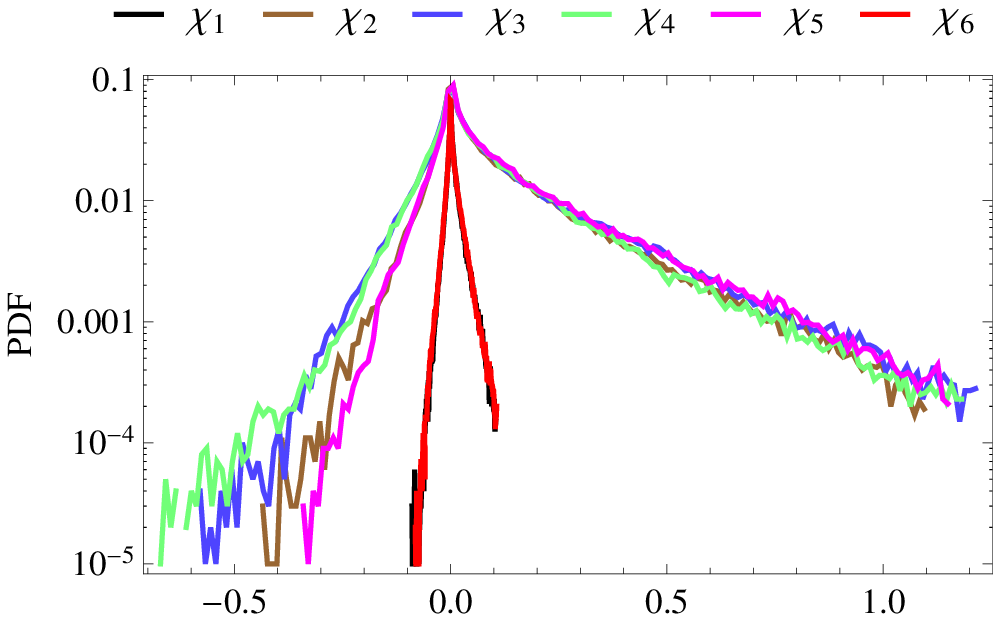}
\hspace{4mm}
\includegraphics[width=0.48\textwidth]{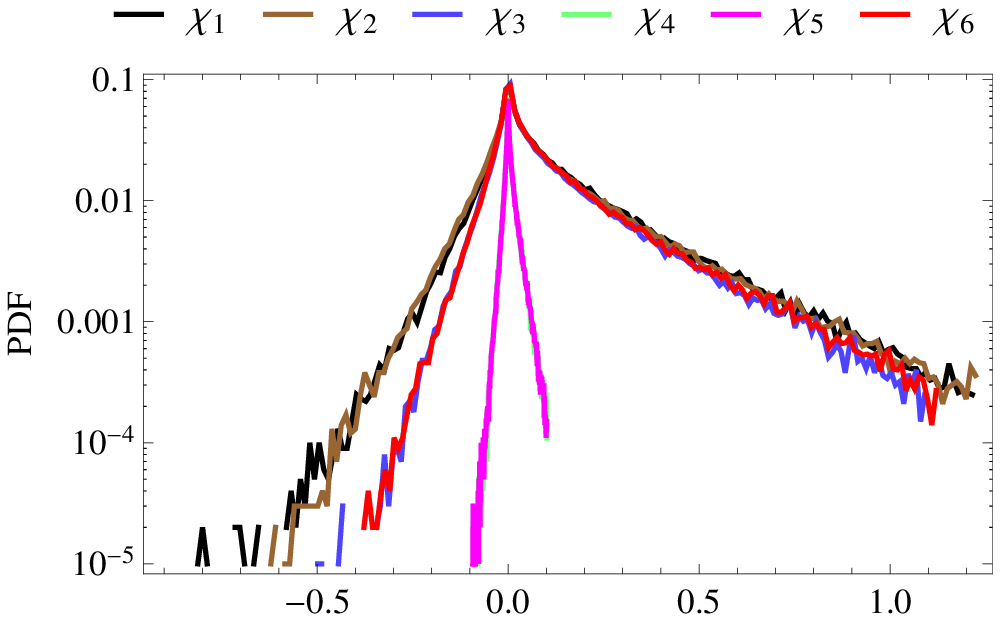}
}
\caption{%
PDF for the six helicity parts same as in Figure~\ref{fig1}:
non-helical anisotropic case for anisotropy in $x$-direction (top
left) and $z$-direction (top right), helical anisotropic case for
anisotropy in $x$-direction (bottom left) and $z$-direction (bottom
right). } \label{fig2}
\end{figure*}

\begin{table*}
\begin{center}
\caption{Summary of the model simulation cases partially presented
on Figure~\ref{fig2}. Note that observationally available parts of
helicity are approximately equal in the case with anisotropy with
respect to $z$ only. They are equal in all helical isotropic cases
while in other anisotropic helical cases they are not equal.}
\begin{tabular}{|c|c|c|c|c|c|c|}
\hline
 \text{case (mean~$\pm$~dispersion)} &
 \text{$\chi_1$} & \text{$\chi_2$} & \text{$\chi_3$} & \text{$\chi_4$} & \text{$\chi_5$} &  \text{$\chi_6$} \\\hline
 \text{Isotopic Non-helical} &
 \text{0.$\pm $2.71} & \text{0.$\pm $2.74} & \text{0.$\pm $2.62} & \text{0.$\pm $2.61} & \text{0.$\pm $2.67} &  \text{0.$\pm $2.67} \\
\hline
 \text{Isotropic Helical} &
 \text{1.56$\pm $3.12} & \text{1.53$\pm $3.12} & \text{1.53$\pm $3.11} & \text{1.55$\pm $3.12} & \text{1.56$\pm $3.12} & \text{1.56$\pm $3.11} \\
\hline \text{Anisotropic-$x$ Non-helical} &
 \text{0.$\pm $0.19} & \text{0.$\pm $1.93} & \text{0.$\pm $2.28} & \text{0.$\pm $2.13} & \text{0.$\pm $1.79} &  \text{0.$\pm $0.19} \\
 \text{Anisotropic-$y$ Non-helical} &
 \text{0.$\pm $1.81} & \text{0.$\pm $0.2} & \text{0.$\pm $0.2} & \text{0.$\pm $1.88} & \text{0.$\pm $2.08} &  \text{0.$\pm $2.14} \\
 \text{Anisotropic-$z$ Non-helical} &
 \text{0.$\pm $2.2} & \text{0.$\pm $2.13} & \text{0.$\pm $1.9} & \text{0.$\pm $0.2} & \text{0.$\pm $0.2} & \text{0.$\pm $1.82} \\
 \hline
 \text{Anisotropic-$x$ Helical} &
 \text{0.05$\pm $0.2} & \text{1.29$\pm $2.24} & \text{1.29$\pm $2.41} & \text{1.36$\pm $2.57} & \text{1.23$\pm $2.04} & \text{0.05$\pm $0.2} \\
 \text{Anisotropic-$y$ Helical} &
 \text{1.29$\pm $2.35} & \text{0.07$\pm $0.22} & \text{0.07$\pm $0.22} & \text{1.38$\pm $2.34} & \text{1.25$\pm $2.44} &  \text{1.29$\pm $2.54} \\
  \text{Anisotropic-$z$ Helical} & \text{1.26$\pm $2.44} & \text{1.26$\pm $2.47} & \text{1.26$\pm $2.19} & \text{0.06$\pm $0.2} & \text{0.06$\pm $0.2} &   \text{1.26$\pm $2.18} \\
\hline
\end{tabular}
\label{tab:rod}
\end{center}
\end{table*}

\begin{figure*} 
 {
{\includegraphics[width=0.32\textwidth]{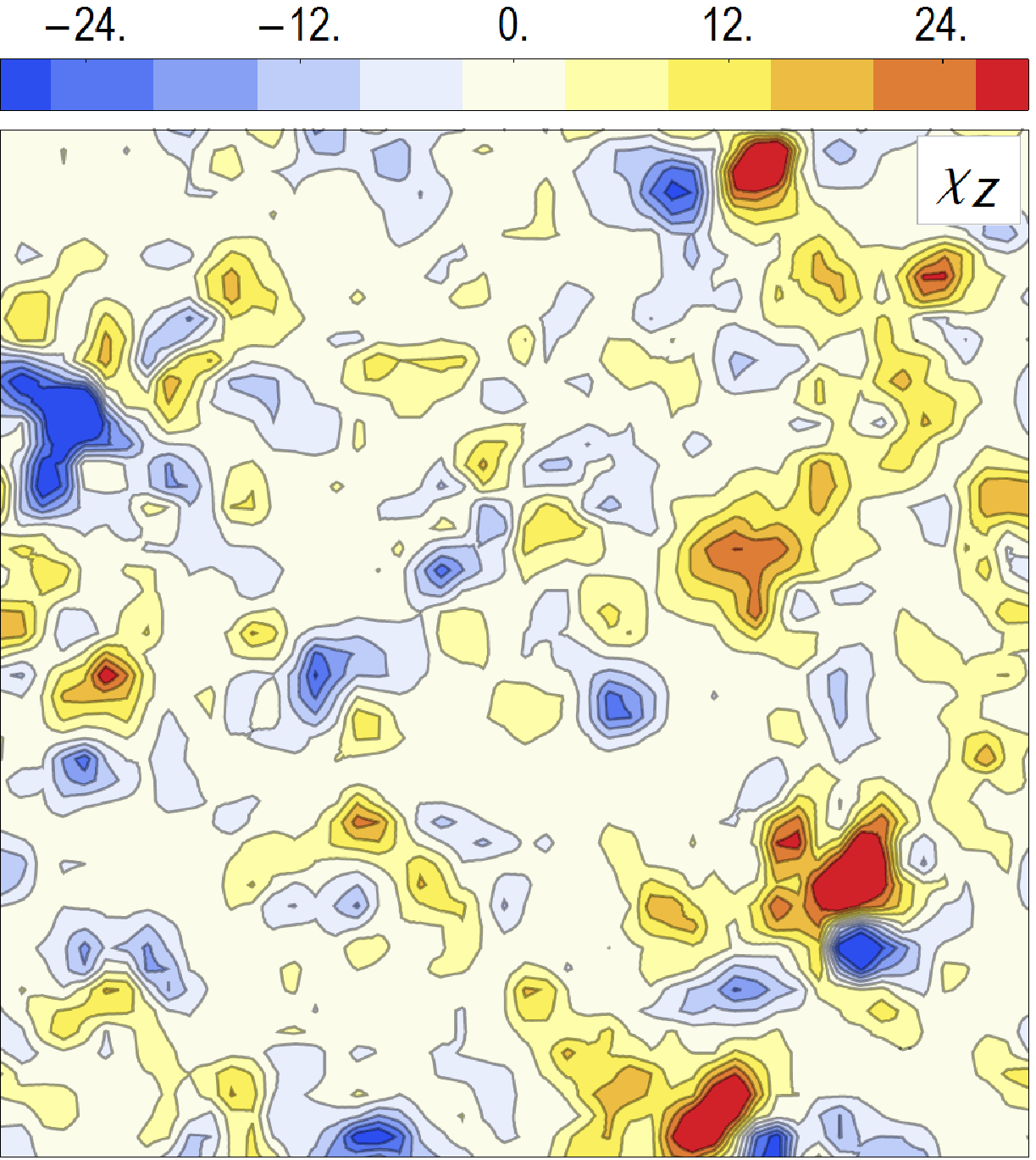}
\includegraphics[width=0.32\textwidth]{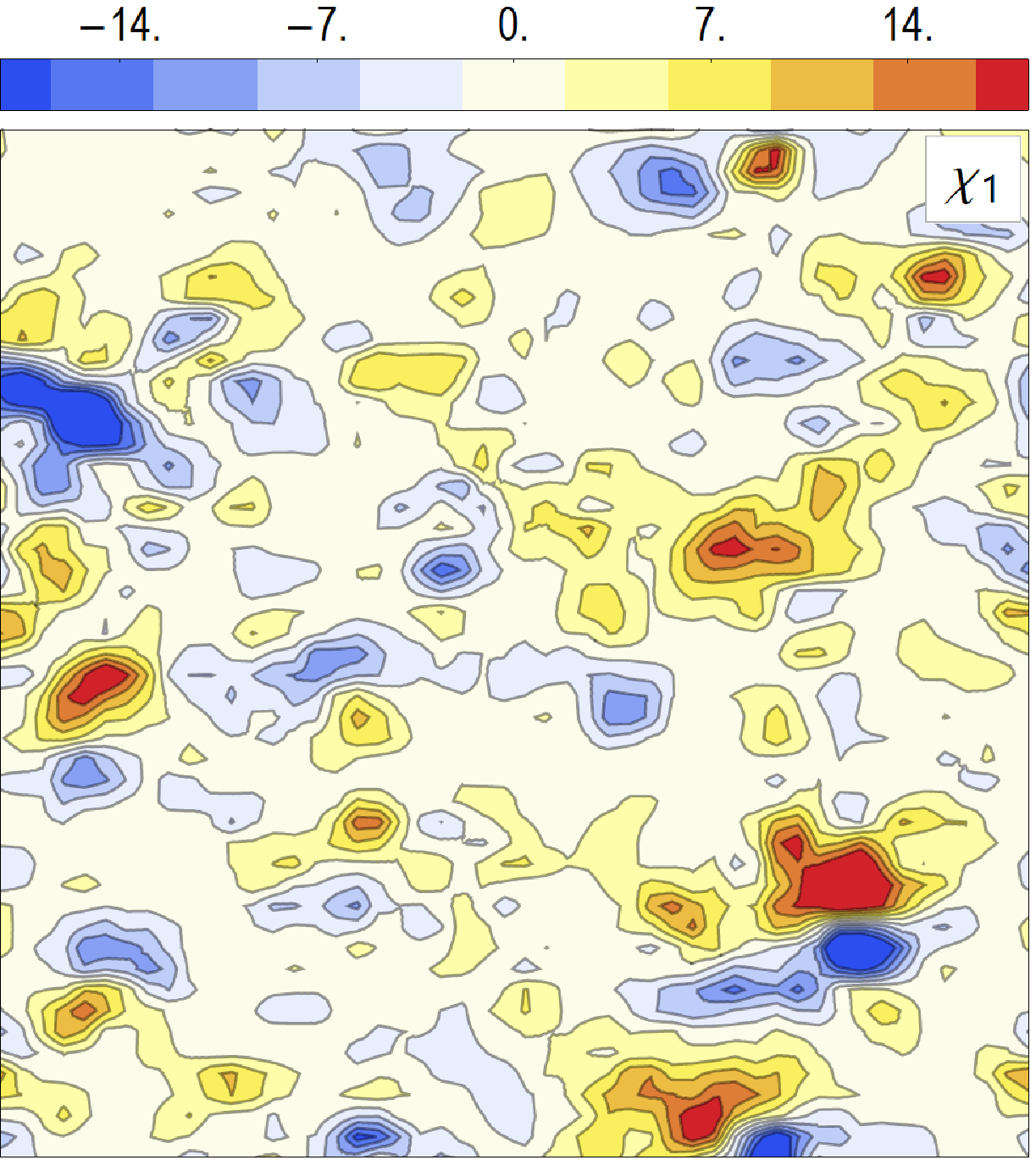}
\includegraphics[width=0.32\textwidth]{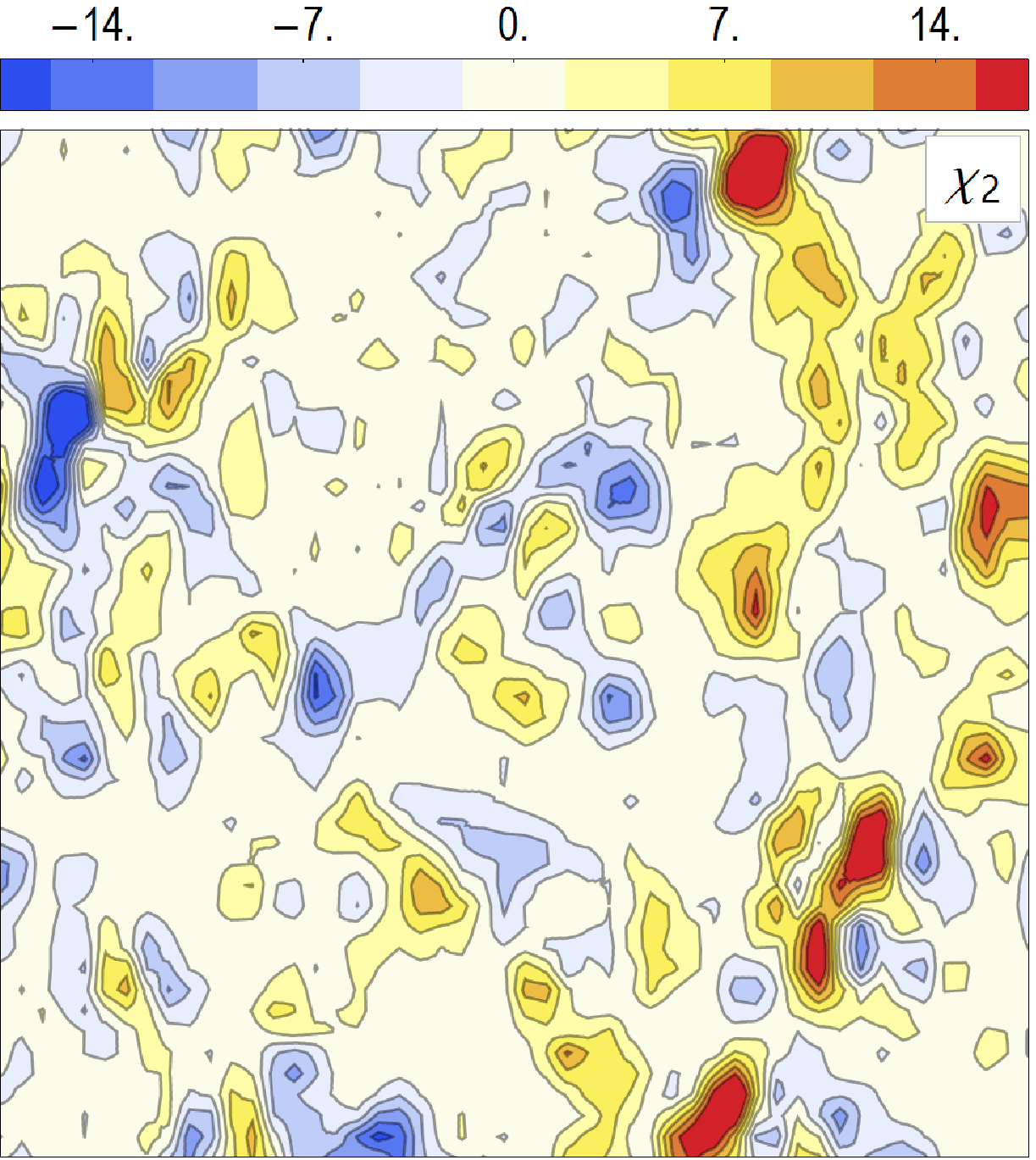}
} \vspace{0.01\textwidth}
{\includegraphics[width=0.32\textwidth]{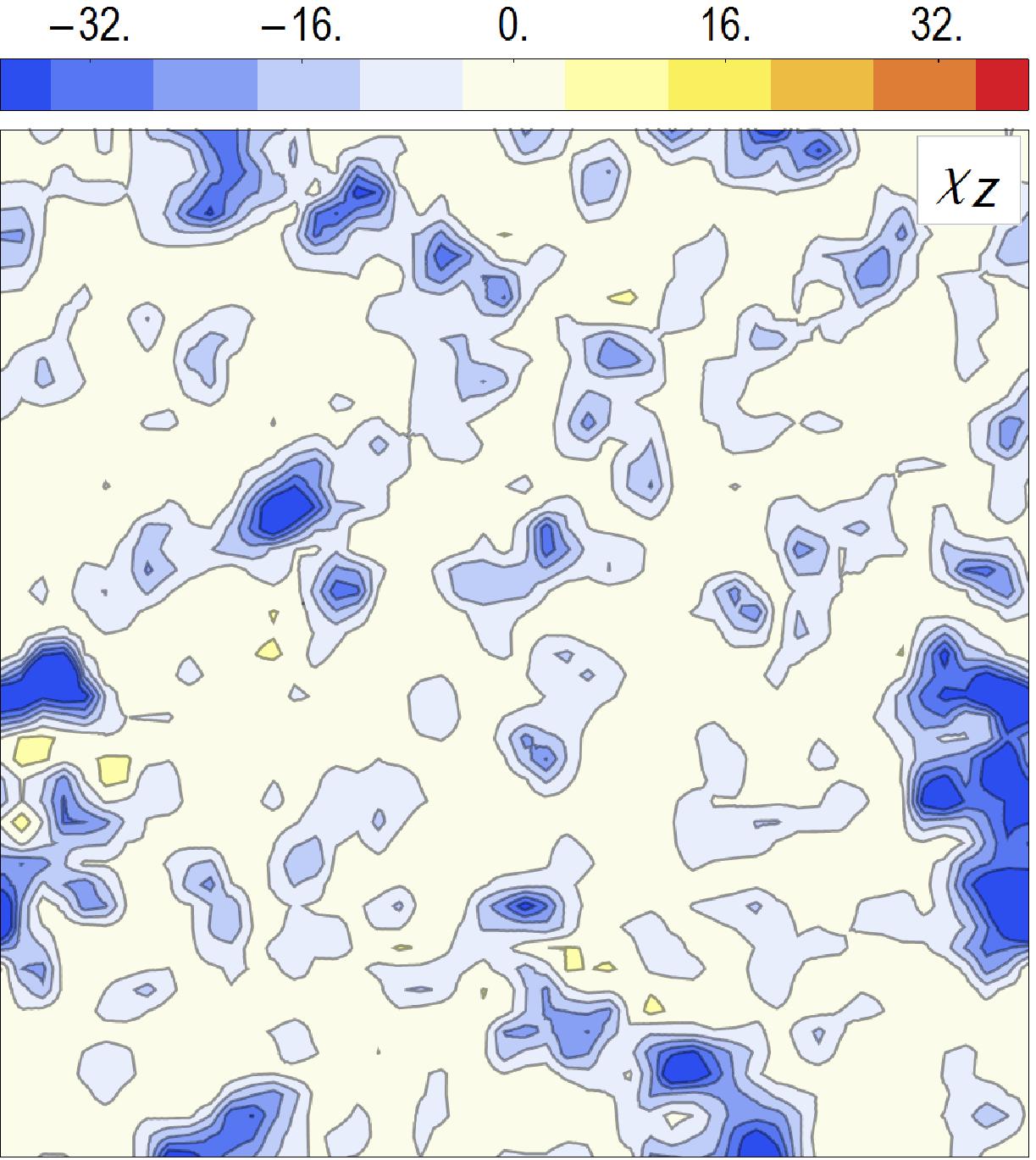}
\includegraphics[width=0.32\textwidth]{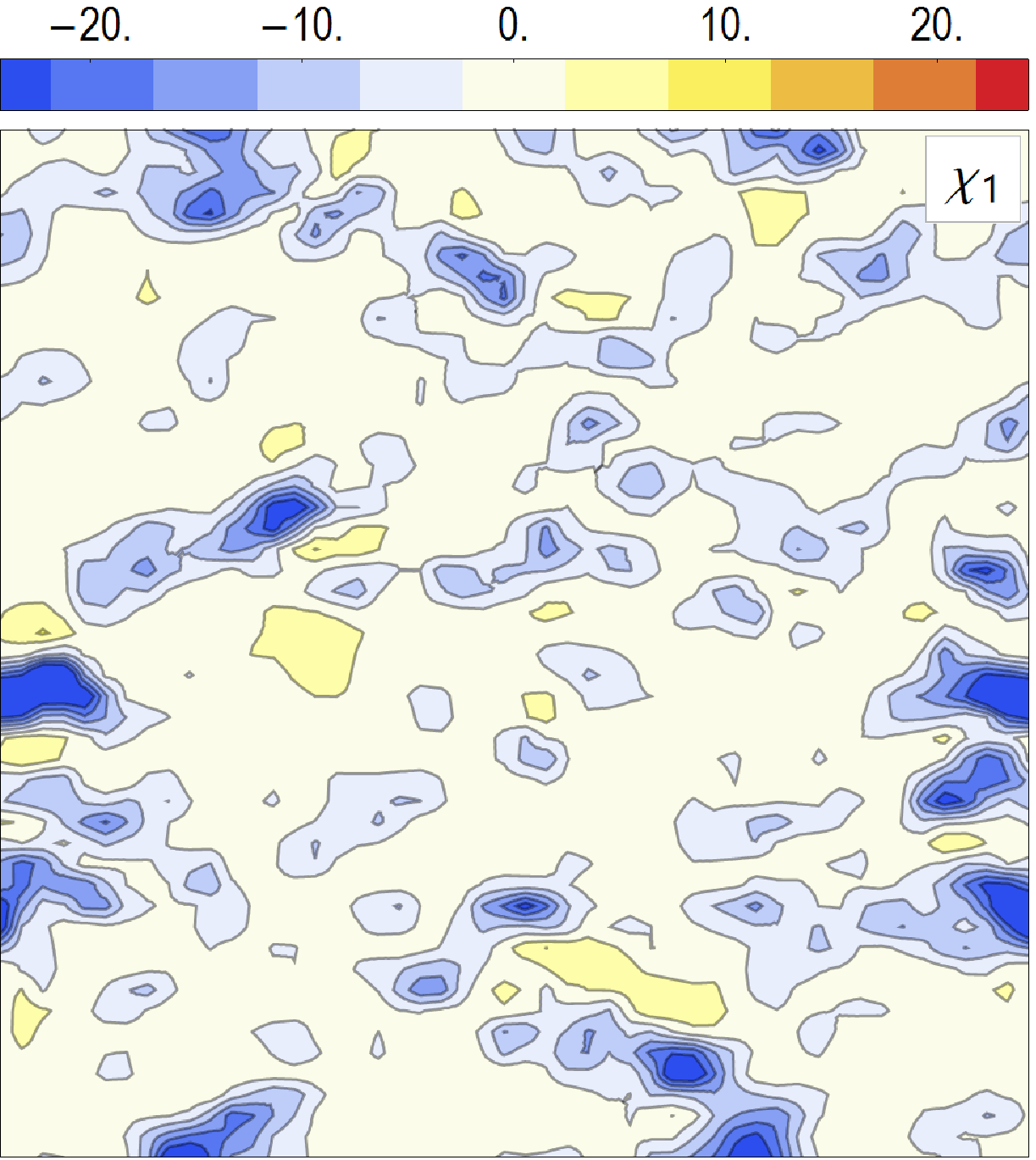}
\includegraphics[width=0.32\textwidth]{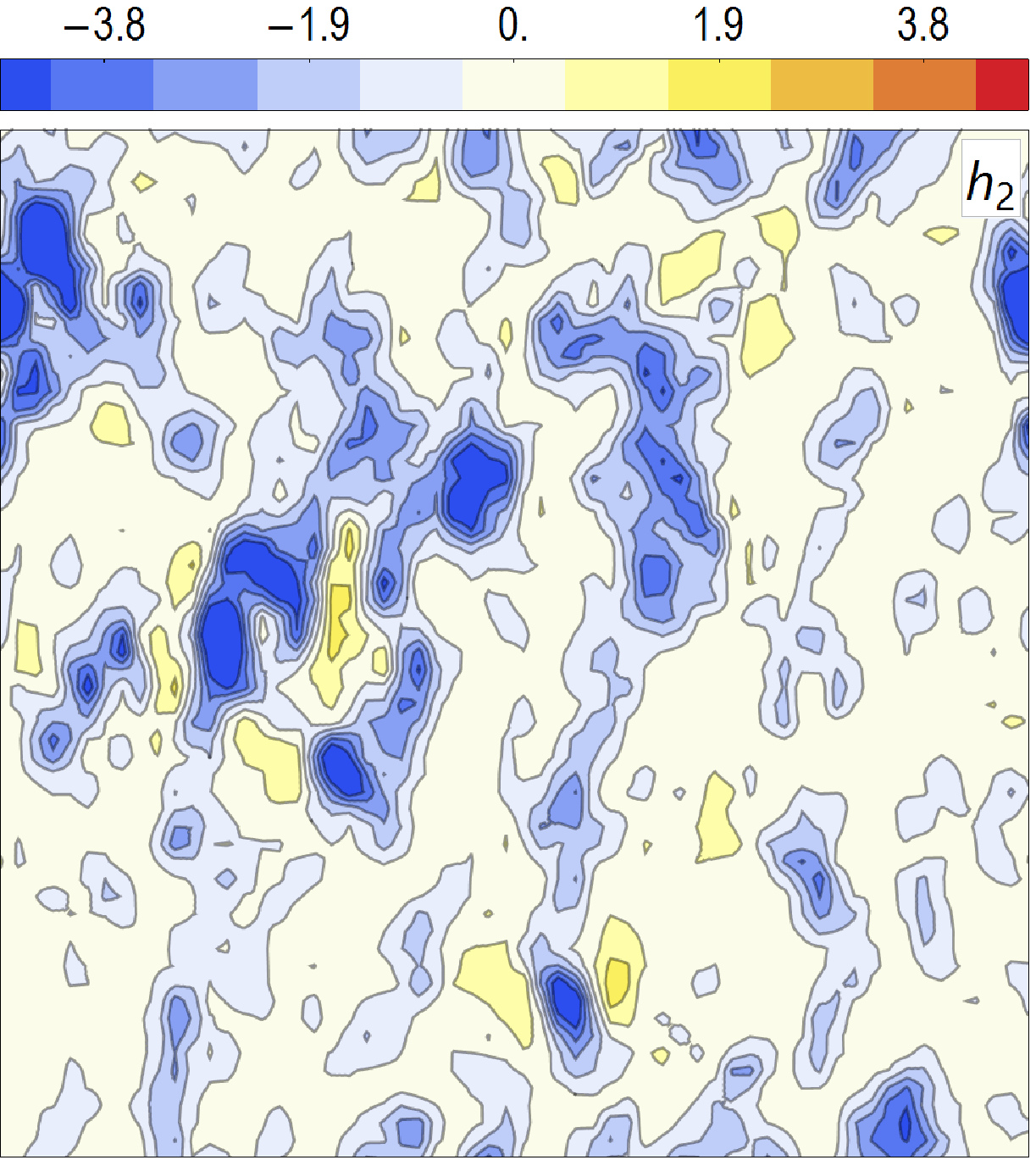}
} \vspace{0.01\textwidth}
{\includegraphics[width=0.32\textwidth]{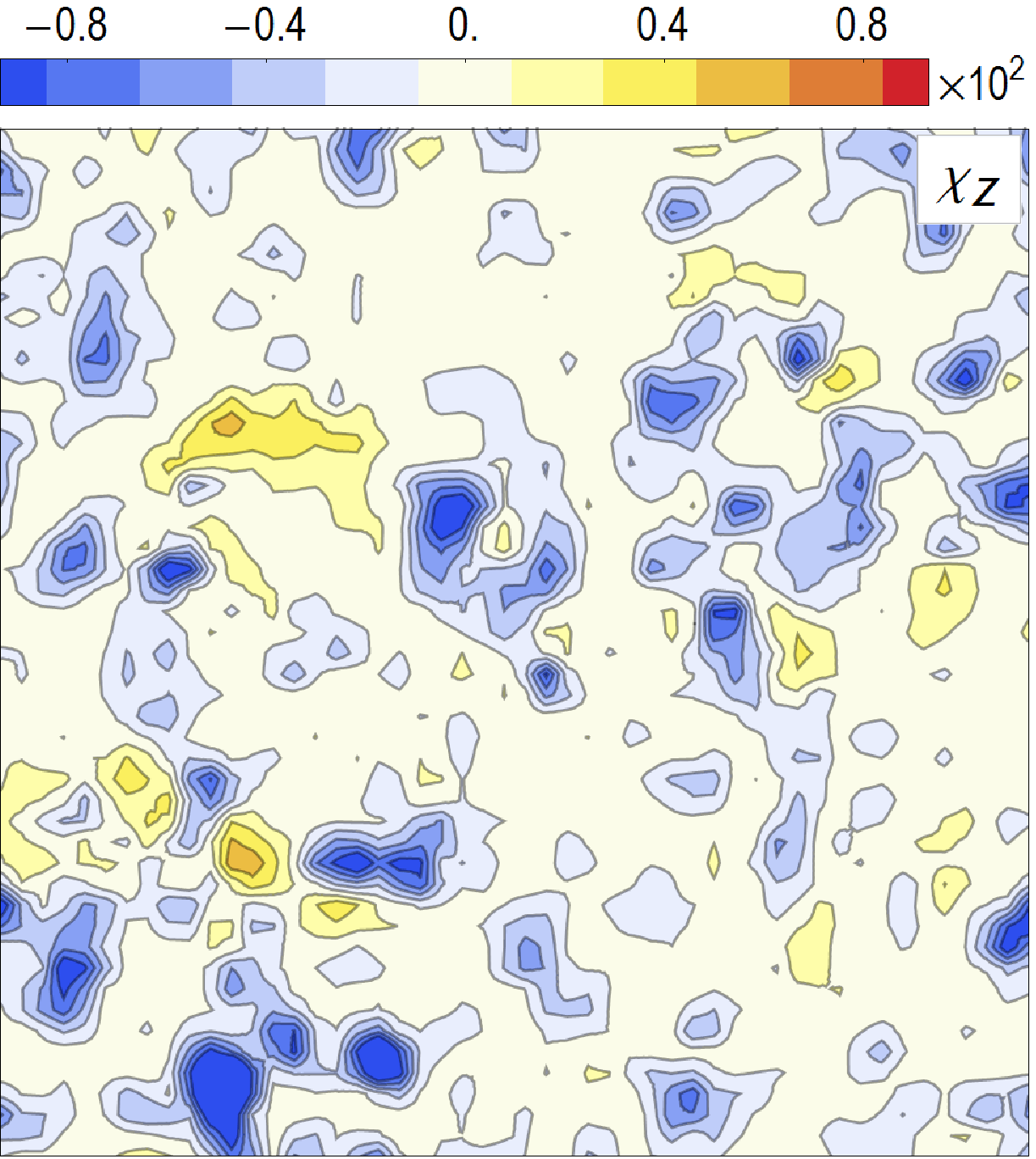}
\includegraphics[width=0.32\textwidth]{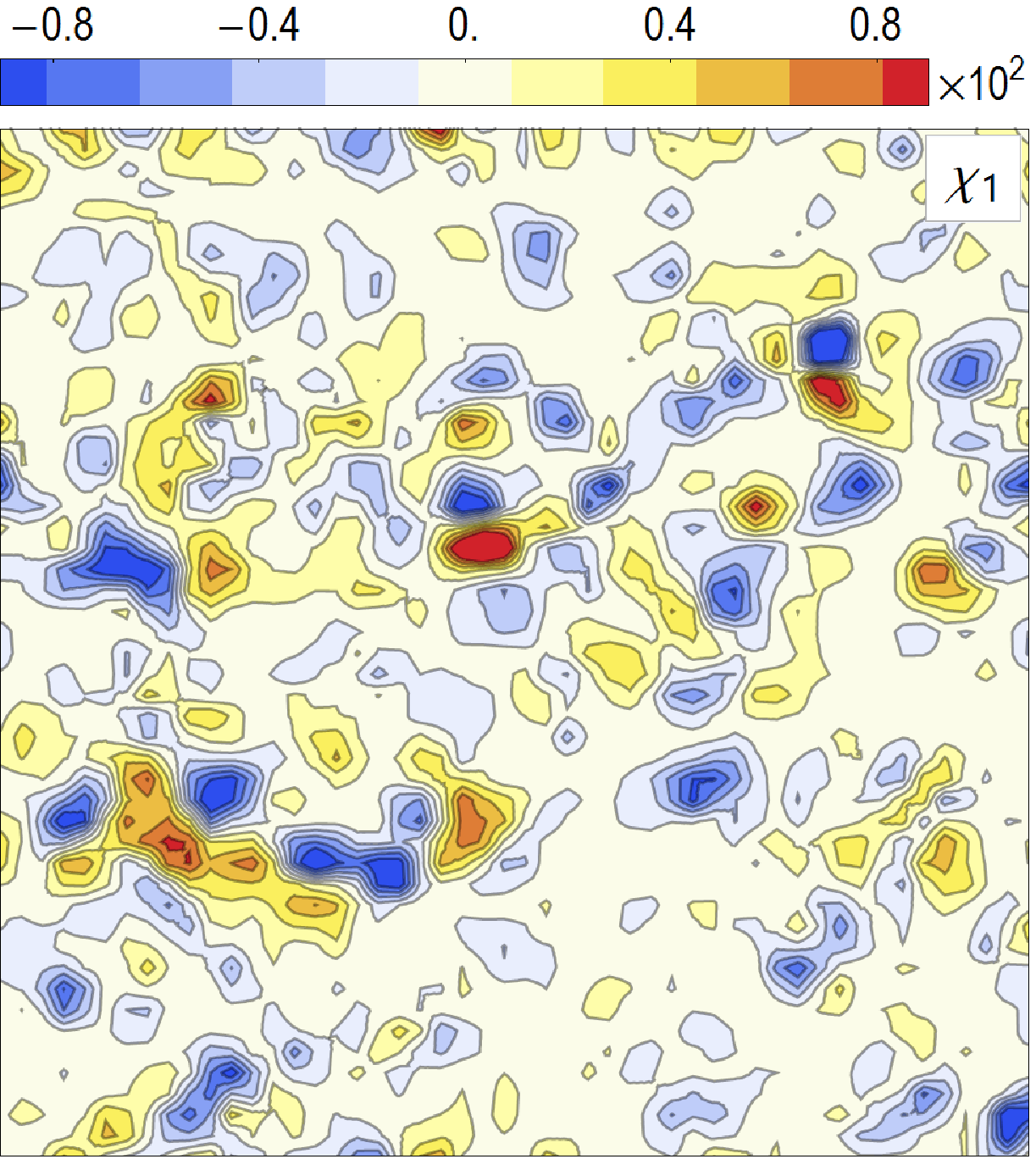}
\includegraphics[width=0.32\textwidth]{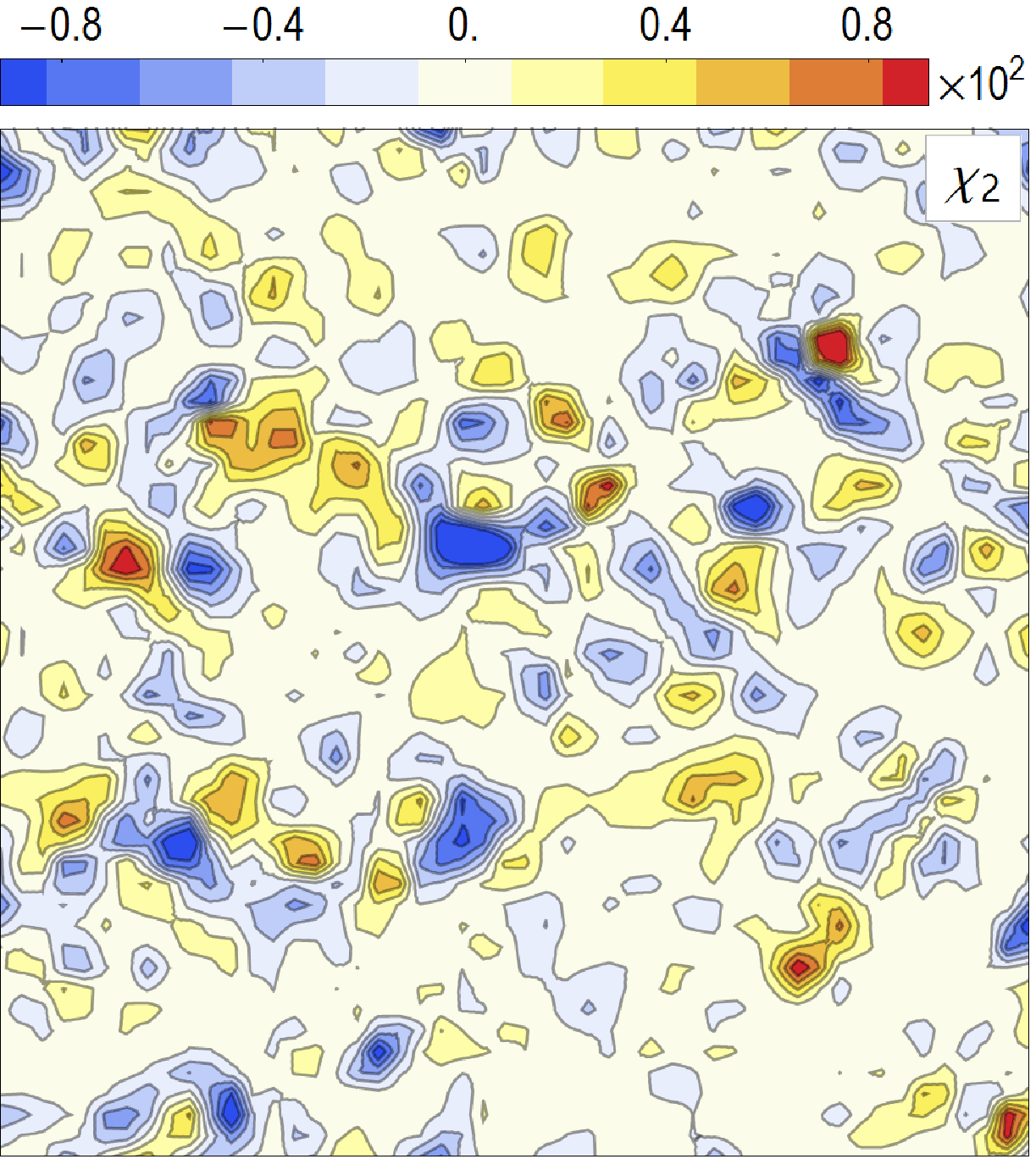}
} }
\caption{%
Distribution of $\chi_z$, $\chi_1$ and $\chi_2$ for three cases:
non-helical magnetic field (top), helical magnetic field (middle),
helical magnetic field plus potential magnetic field (bottom). }
\label{fig3}
\end{figure*}

To prescribe a quasi-random magnetic field $\vec{B}$ with vanishing mean value in a periodic box, we use a Fourier expansion in modes with randomly chosen directions of wave vectors $\vec{k}$ but with amplitudes adjusted to reproduce any desired energy spectrum:
\begin{equation}
\vec{B}(\vec{x})=\frac{1}{(2\pi)^{3/2}}\int \hat{\vec{B}} (\vec{k})
e^{\mathrm{i} \vec{k}\cdot\vec{x}}\, \dd^3\vec{k}, \label{ft}
\end{equation}
where $ \hat{\vec{B}}$ is the Fourier transform of $\vec{B}$. The
corresponding magnetic energy spectrum is given by
\begin{equation}
M(k)=\int_{|\vec{k}'|=k} |\hat{\vec{B}}(\vec{k}')|^2\,
\dd^3\vec{k}', \label{spec}
\end{equation}
where the integral is taken over the spherical surface of radius $k$
in $k$-space. In the isotropic case, $M(k)=4\pi k^2
|\hat{\vec{B}}(k)|^2$. In order to ensure periodicity within a
computational box of size $L$, as required for the discrete Fourier
transformation, the components of the wave vectors are restricted to
be integer multiples of $2\pi/L$.

\begin{figure*}
{
\includegraphics[width=1.0\textwidth]{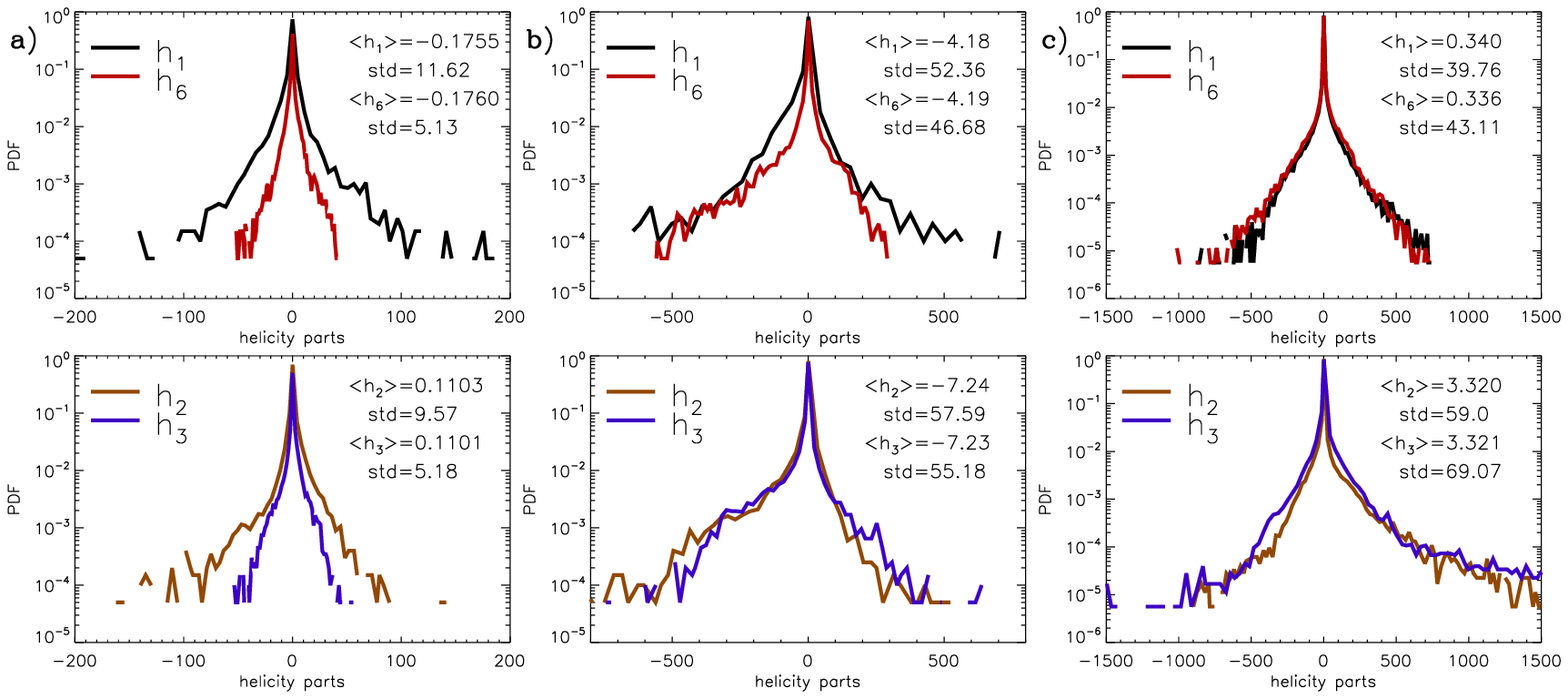}
}
\caption{%
The examples of PDFs for the four observable helicity parts: a)
column is for NOAA 8898; b) column is for NOAA 6659; c) column is
for NOAA 11158. The upper (bottom) panels are the PDFs for helicity
parts $h_1$ and $h_6$ ($h_2$ and $h_3$). The mean values and the
standard deviations of helicity parts are given on the top right
corner of each panel.  } \label{fig5}
\end{figure*}

\begin{figure*}
{\includegraphics[width=1.0\textwidth]{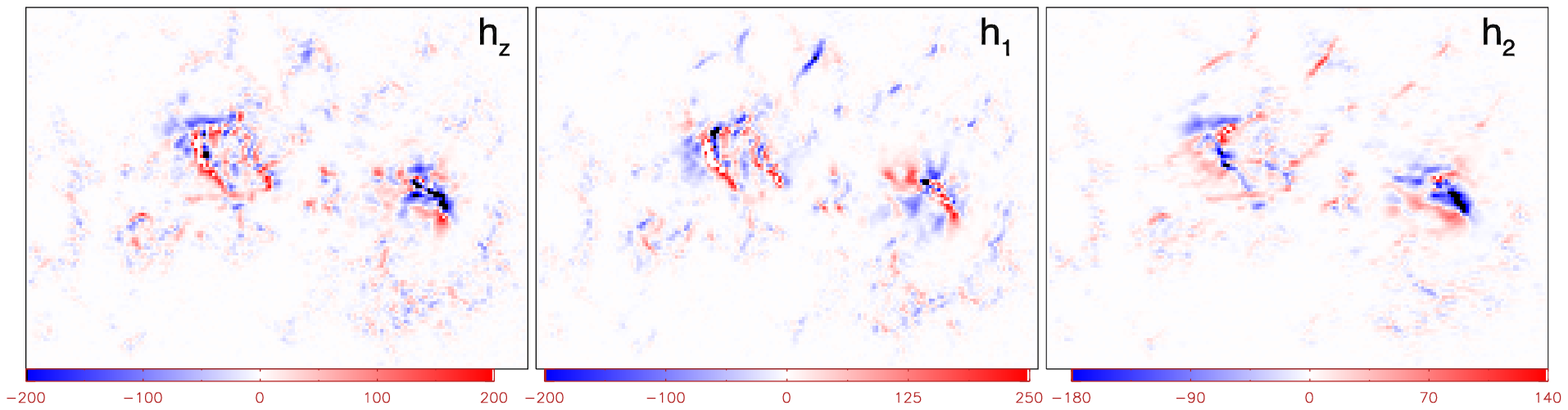}
\includegraphics[width=1.0\textwidth]{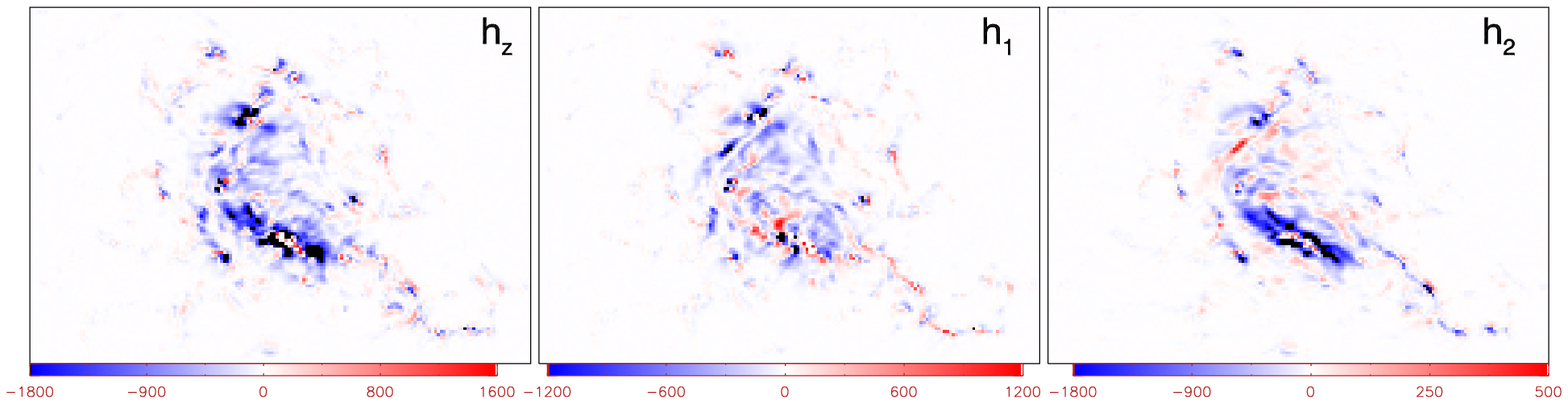}
\includegraphics[width=1.0\textwidth]{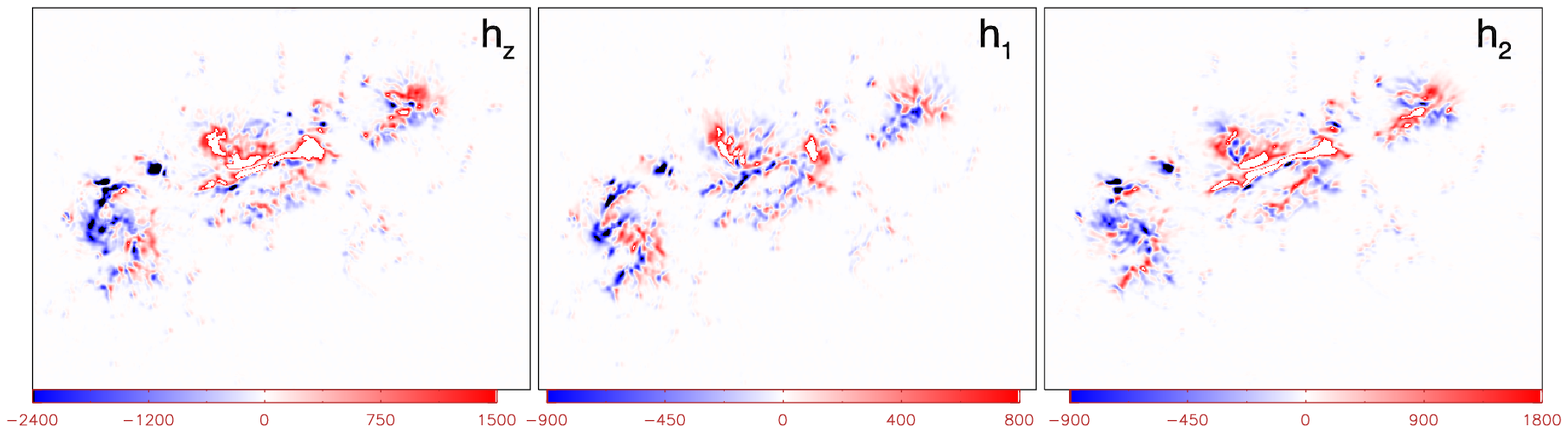}
}
\caption{%
Distribution of $h_z$, $h_1$ and $h_2$ for the three examples:
Nearly potential magnetic field in NOAA 8898 (top), helical magnetic
field in NOAA 6659 (middle), relatively high special resolution
magnetic field in NOAA 11158. The unit of helicity is
$10^{-3}G^{2}/m$.} \label{fig6}
\end{figure*}

A solenoidal vector field $\vec{B}$, i.e., that having
$\vec{k}\cdot\hat{\vec{B}}(\vec{k})=0$, is specified by
\[
\hat{\vec{B}}(\vec{k})=\frac{\vec{k}\times\vec{X}}{|\vec{k}\times\vec{X}|}k^{-1}\sqrt{M(k)}.
\]
Random choice a complex vector  $\vec{X}$ implies zero net current helicity of $\vec{B}$.
We consider a magnetic energy spectrum represented by two power-law ranges,
\begin{equation}
M(k)=M_0
\left\{
\begin{array}{ll}
(k/k_0)^{s_0}    &\mbox{for } k < k_0,\\
(k/k_0)^{-s_1}    &\mbox{for } k\geq k_0,
\end{array}
\right.
\label{Mss}
\end{equation}
with $s_0 > 0$, $s_1 > 0$ and $M_0 =1$, where $k_0=6$ is the
energy-range wave-number. We use $s_1=5/3$ as in Kolmogorov's
spectrum \citep{Ko41} and $s_0=2$ as in \cite{Christensson2001}.

{\bf The helical $\vec{B}$ can be obtained with choice $\vec{X}$ as
\begin{equation}
\vec{X}=\vec{Y} \pm \mathrm{i} |\vec{Y}|
\frac{\vec{k}\times\vec{Y}}{|\vec{k}\times\vec{Y}|}, \label{Mss1}
\end{equation}
where the sign defines the sign of current helicity and
$\vec{Y}(\vec{k})=\vec{Y}(-\vec{k})$ is a random real vector.
Condition (\ref{Mss1}) implies
$\hat{\vec{B}}(\vec{k})=\hat{\vec{B}}^*(-\vec{k})$.}

The probability density functions (PDFs) calculated from 2D
distributions [$(x,y)$-plane of 3D simulated cube] are shown in Fig.
\ref{fig1}. First of all we note that isotropy means similarity of
distributions of $\chi_1$, $\chi_2$, $\chi_3$ and $\chi_6$.
Secondly, non-zero helicity leads to asymmetry of PDFs. Furthermore,
we can produce corresponding statistically anisotropic fields for
non-helical and helical cases.

Anisotropic case is simulated by the additional factor in $M(k)$. One can take
\begin{equation}
M(k)= k_x^{-2} M_0
\left\{
\begin{array}{ll}
(k/k_0)^{s_0}    &\mbox{for } k < k_0,\\
(k/k_0)^{-s_1}    &\mbox{for } k\geq k_0,
\end{array}
\right.
\label{Mss2}
\end{equation}
The corresponding PDFs are shown in Figure~\ref{fig2}. From
Figure~\ref{fig2} one can see that anisotropy in different direction
affects the PDFs of different parts of helicity in a different way.
Anisotropy in $x$-direction on the left two panels leads to the
parts of helicity $\chi_1$ and $\chi_6$ containing derivatives in
this direction to be distributed with much lower dispersion. One can
see the same for anisotropy in $z$-direction in the right panels for
$\chi_4$ and $\chi_5$. Each of the two quantities in these pairs are
distributed statistically similar. For the remaining parts of
helicity, in the helical case (bottom panels), we can see that the
left and right tails of the PDFs have different spreads and in
particular the left tails are more inclined. Furthermore, we see
that these four remaining parts of helicity group in two
statistically similar pairs, with respect to the component of the
magnetic field which enters into each of the parts, namely $B_x$ for
$\chi_3$ and $\chi_4$ in case of anisotropy in $x-$~direction and
$B_Z$ for $\chi_1$ and $\chi_2$ in case of anisotropy in
$z-$direction.

The results for the cases shown in Figure~\ref{fig2} are summarized
in Table~\ref{tab:rod}. One can see that the mean values of the six
helicity parts are approximately equal in the isotropic helical
case. In the anisotropic helical cases, in accord with the
integration by parts formula (\ref{int_by_parts}), the mean values
of helicity parts are equal in pairs: $ \langle \chi_1 \rangle
\approx \langle \chi_6 \rangle $ and $ \langle \chi_2 \rangle
\approx \langle \chi_3 \rangle $. The other two parts $ \langle
\chi_4 \rangle $ and $ \langle \chi_5 \rangle $ are not subject of
the formula (\ref{int_by_parts}) as they contain derivatives with
respect to $z$ while averaging involves differentiation over the
image plane $(x,y)$, and they may generally be different, as it is
in the anisotropic helical case.

Now let us consider how the sign of the parts $\chi_1$, $\chi_2$,
$\chi_3$ and $\chi_6$ can locally represent the sign of the total
helicity. One effect can be the contribution from the potential
magnetic field. We simulate distributions of $\chi_z$, $\chi_1$ and
$\chi_2$ for three cases: non-helical magnetic field, helical
magnetic field and helical magnetic field plus potential magnetic
field (see Figure~\ref{fig3} from top to bottom). As expected for
the purely non-helical magnetic field, the all three maps possess
the same kind of patterns with alternating sign. For the helical
case we also have the same kind of patterns but with dominating sign
of helicity. Additional contribution from the potential magnetic
field does not change much the total helicity $ \chi_z$ but its
parts $\chi_1$ and $\chi_2$ have strongly alternating pattern unlike
for the case of non-helical magnetic field. We can see that
assigning these two parts the opposite signs in the alternating
pattern may merely cancel each other.

Therefore, we have seen that {\bf the four parts} of helicity that
have observational interest are close in pairs, but between the
pairs there may be a significant difference in their distributions
and the integral values. We have also noted now the specific
contribution of one and the other parts may partially cancel each
other in the overall helicity.

\section{Observational Results}

Now we compare theoretical predictions with results of observational
data analysis. Figures~\ref{fig5} show PDFs for the four
observationally available parts of helicity $h_1$, $h_2$, $h_3$,
$h_6$ computed for all pixels in the magnetograms of the three
active regions shown in Figures~\ref{fig4}. One can see that the
mean values of distribution for pair ($h_1$, $h_6$), and pair
($h_2$, $h_3$) are very close to each other with accuracy of a few
per cent,  but the difference between the mean values of $h_1$ and
$h_2$ ($h_3$ and $h_6$) is large although the distribution pattern
is similar. This observational result is similar with the
anisotropic case simulation in Figure~\ref{fig2} although some
difference is found. This discrepancy can be attributed to the
complexity of the observation. We also show the distribution map for
$h_z$, $h_1$ and $h_2$ in Figure~\ref{fig6}. We can see that there
is a significant difference in the distribution of $h_1$ and $h_2$,
but their sum is nearly equal to $h_z$.

We have analyzed similar distributions for various vector
magnetograms obtained at Huairou Solar Observing Station (HSOS)
available for date in cycles 22 and 23 as well as a recent vector
magnetograms observed by HMI/SDO in order to learn that the
properties above for this PDF is quite generic for the magnetograms
under consideration.

Now from the study of distribution of the parts of helicity by
pixels over a given magnetogram of an active region we move towards
study the distribution of the mean values of these quantities for
6629 magnetograms observed by HSOS from 1988 to 2005.

\begin{figure}
\centering \resizebox{0.48\textwidth}{!}
{\includegraphics[]{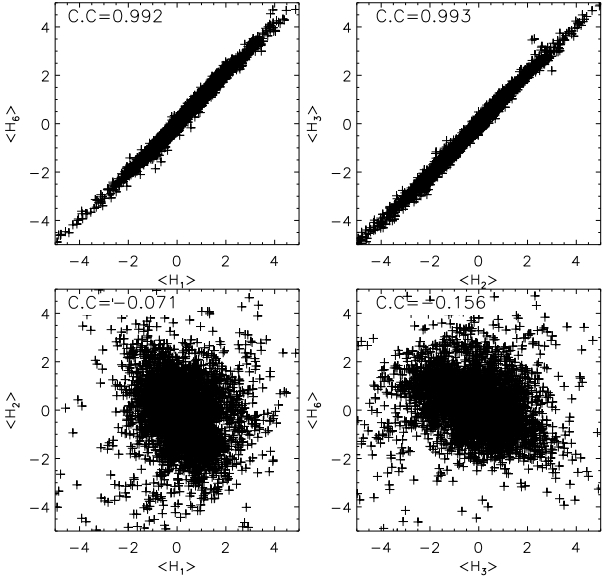} }
\caption{%
Scatter plots key helicity parts over the dataset 1988-2005. each
point represents average value of helicity parts for one
magnetogram. The unit of helicity is $10^{-3}G^{2}/m$.} \label{fig7}
\end{figure}

Figure~\ref{fig7} shows scatter plots for helicity parts for our
sample. One can see that correlation between $ H_1 $ and $ H_6 $ ($
H_2 $ and $ H_3 $) is very high while the one between $  H_1 $ and $
H_2 $ ($ H_3 $ and $ H_6 $) is low. This means that average with
signed flux $b_z$ weighted by $\displaystyle{-\frac{\partial
b_x}{\partial y}}$ is systematically different from the one weighted
by $\displaystyle{-\frac{\partial b_y}{\partial x}}$ in a
magnetogram. These confirm that the properties which we see above
for the PDFs of helicity parts in a magnetogram: robustness of
numerical scheme for the use of integration by parts, as well as
absence of isotropy. Comparing with our theoretical simulation in
the section above, this statistical observational result corresponds
to anisotropic helical case while anisotropy is either by $x$ or
$y$, or both. This is not the case of purely anisotropic by only
vertical $z$-direction (convective stratified turbulence). We also
studied the sign agreement between helicity parts. We found that
95.5\% (95.9\%) of $ H_1 $ and $ H_6 $ ($ H_2 $ and $ H_3 $) among
6629 vector magnetograms agree in their signs, while for $ H_1 $ and
$ H_2 $ ($ H_3 $ and $ H_6 $) the percent of agreement is much lower
which is 38.0\% (34.7\%). We checked the vector magnetograms whose $
H_1 $ and $ H_6 $ ($ H_2 $ and $ H_3 $) have different sign and
found that the magnetic fields are not weak at the boundary for some
of those magnetograms. On the other hand, the values of helicity
parts inferred from those magnetograms are often smaller by one or
two orders of magnitude than for the typical magnetograms. Those
values are too small to be estimated accurately.

\begin{figure*} 
\centering \resizebox{1.0\textwidth}{!}
{\includegraphics[]{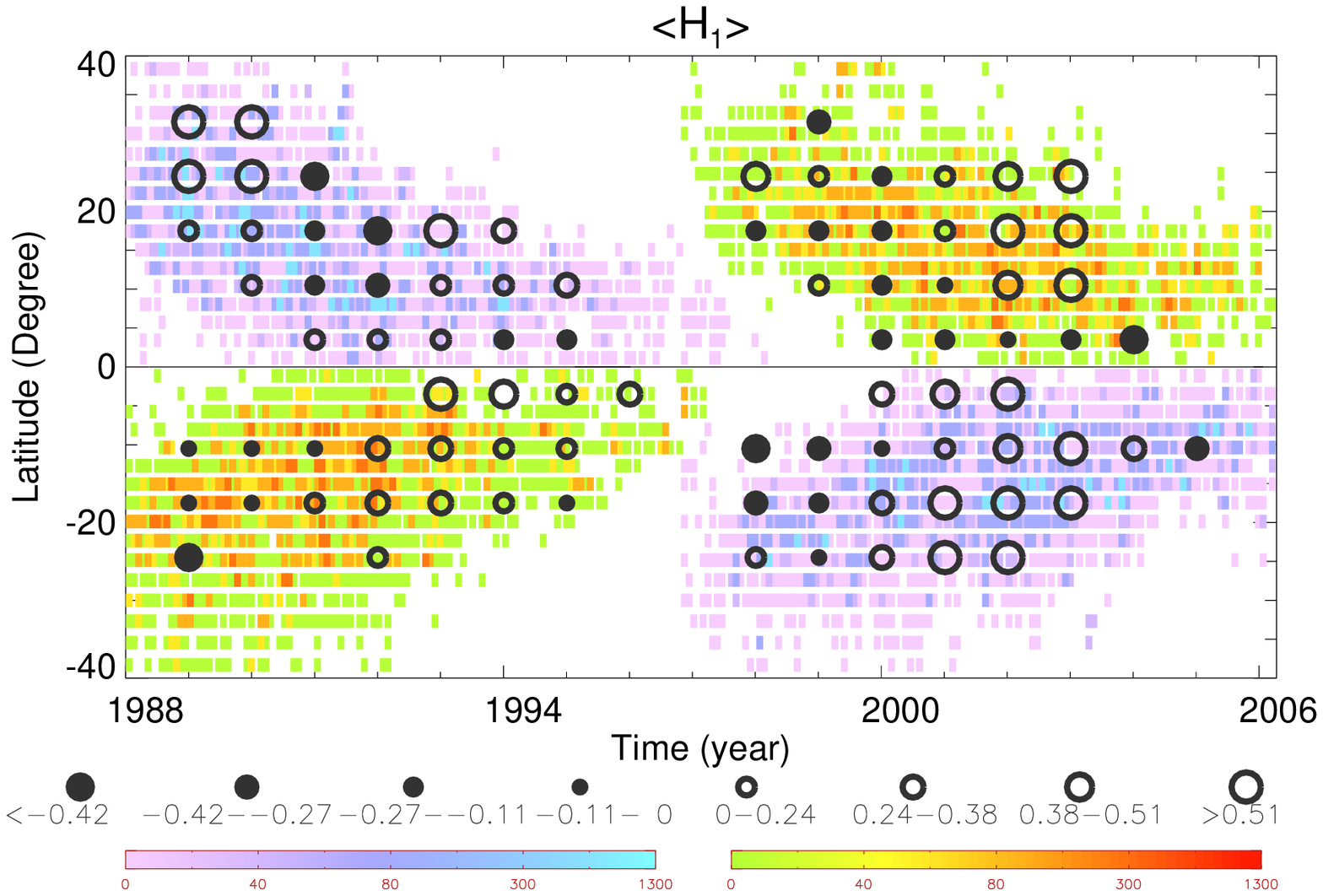}{\includegraphics[]{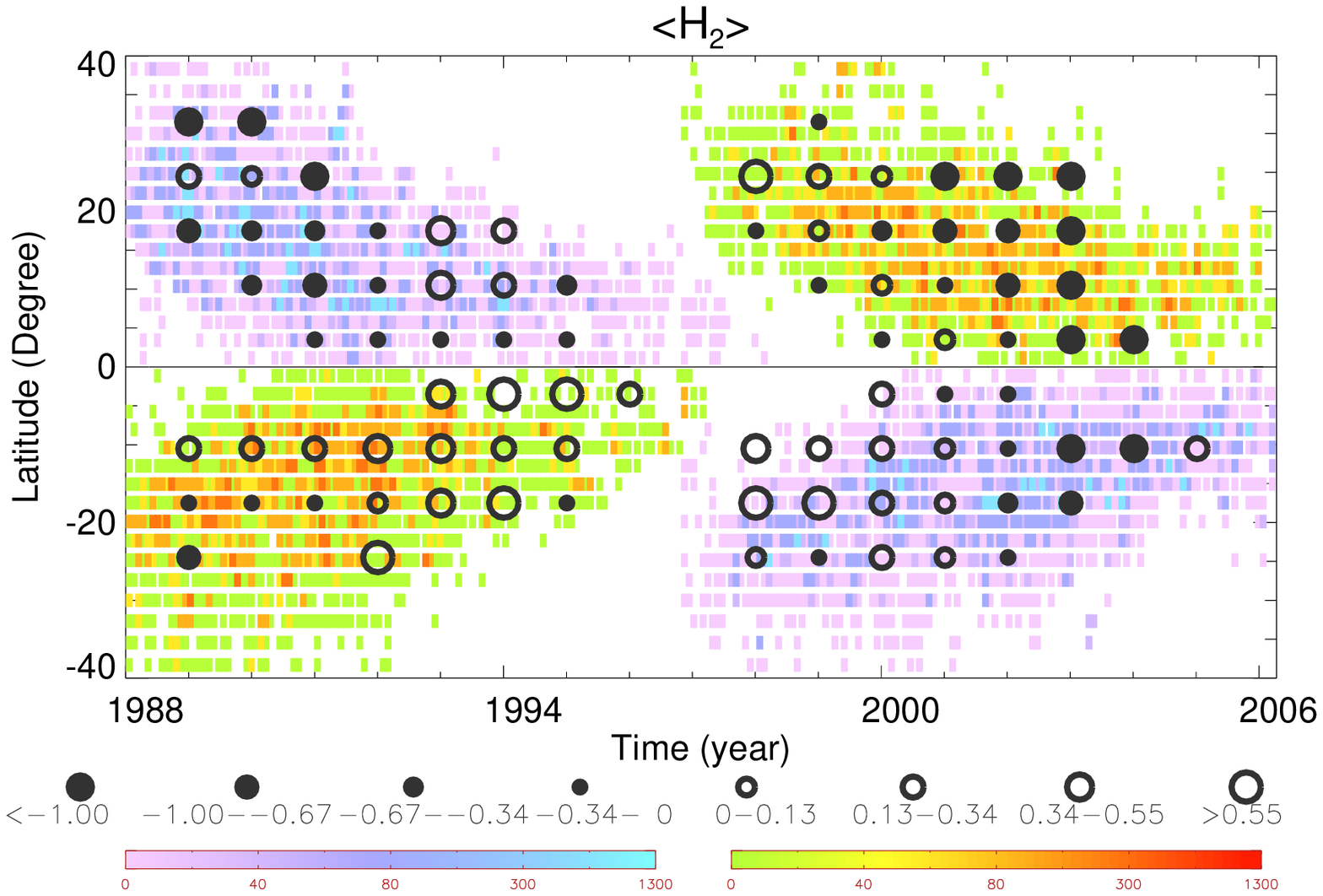}}}
\resizebox{1.0\textwidth}{!}
{\includegraphics[]{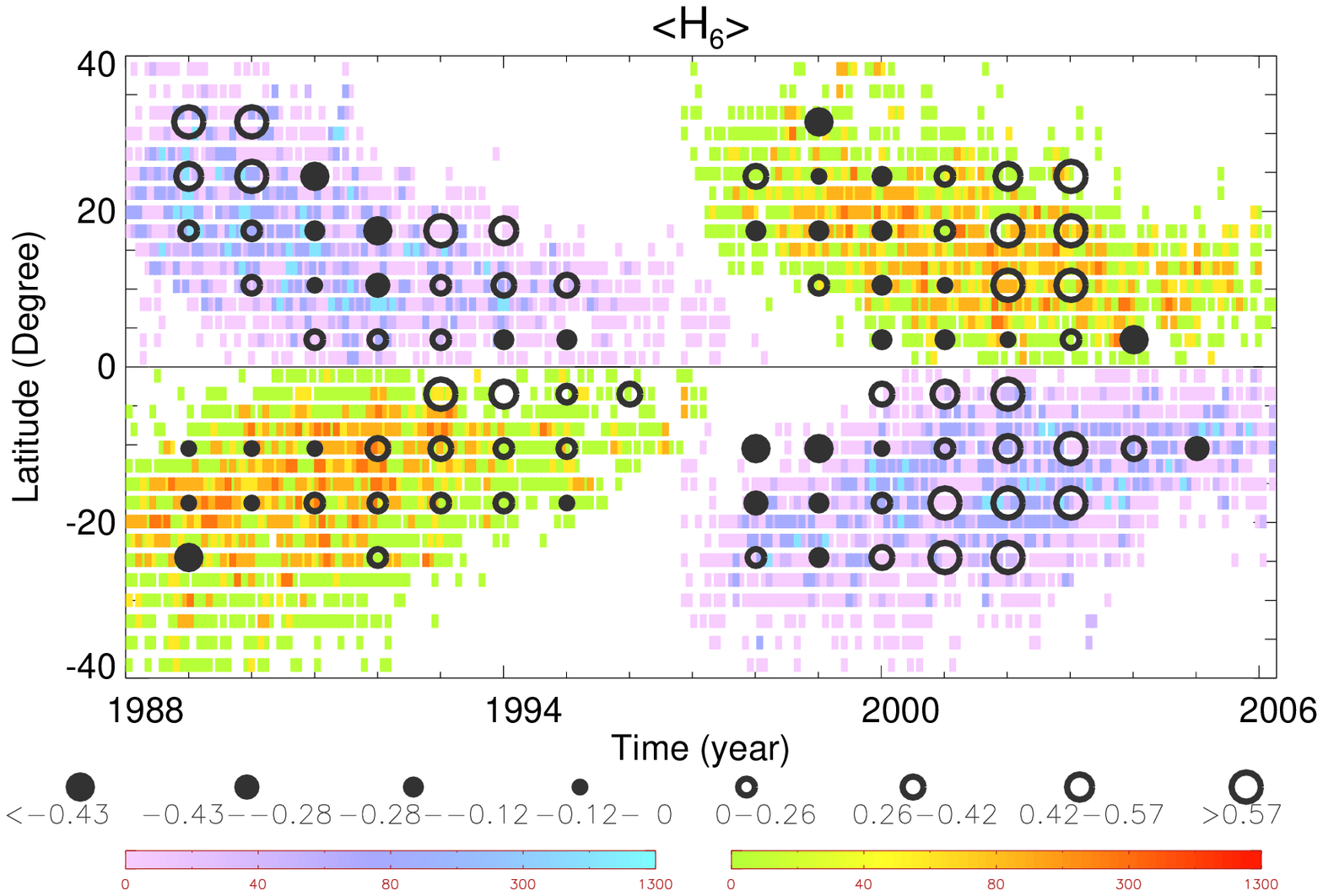}{\includegraphics[]{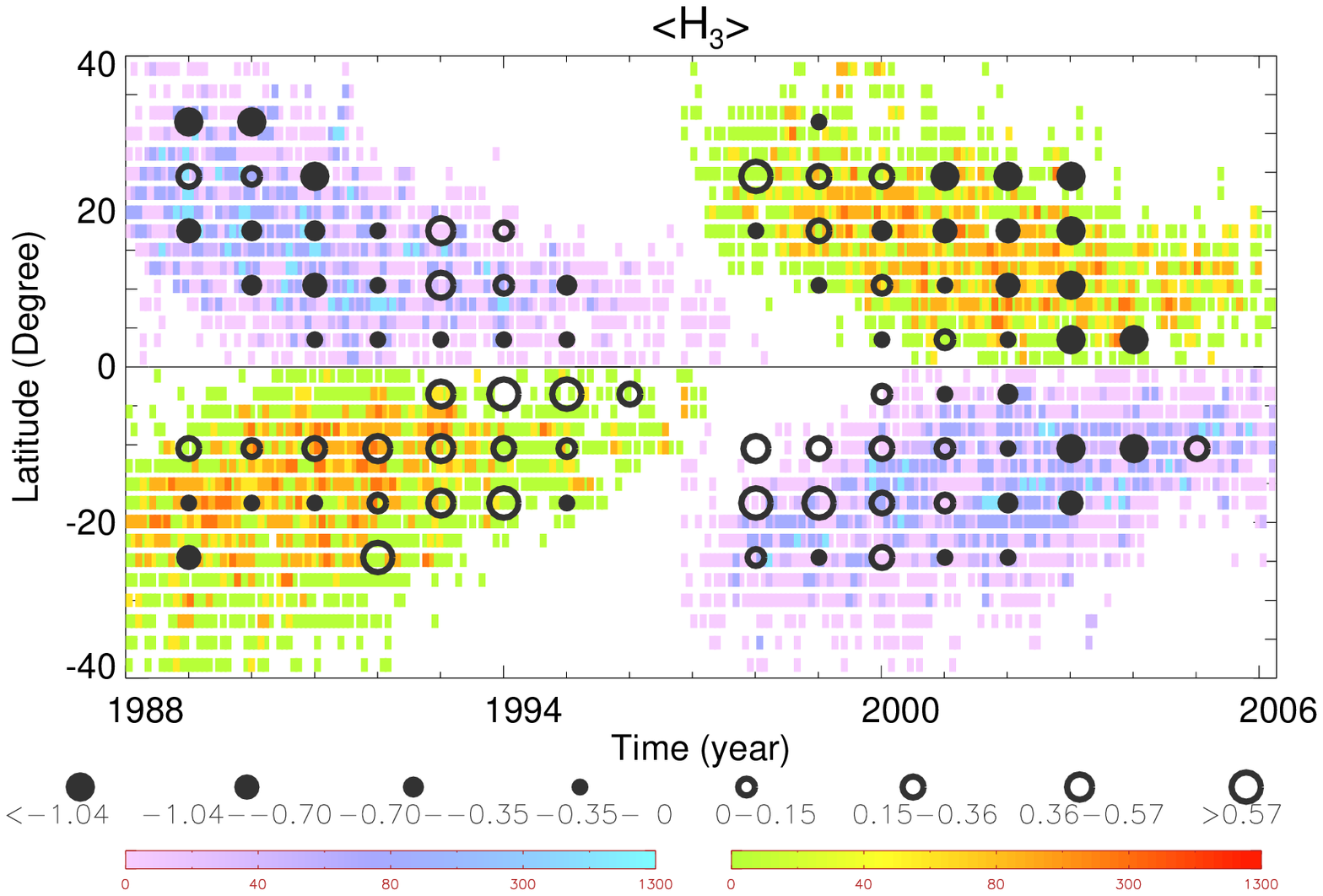}}}
\caption{%
Evolution of four helicity parts with the solar cycle overlaid with
sunspot density (color). The vertical axis gives the latitude and
the horizontal axis gives the time in years. The circle size gives
the magnitude of helicity as averaged over two-year running windows
over latitudinal bins of $7^{\circ}$ wide. The unit of helicity is
$10^{-3}G^{2}/m$.} \label{4butt}
\end{figure*}

Let us study how the local anisotropy of current helicity varies
with the solar cycle. From Figure~{\ref{4butt}},  we can see that
the latitudinal structure and  variation with the solar cycle of $
\langle H_1 \rangle $ and $ \langle H_6 \rangle $ ( $ \langle H_2
\rangle $ and $ \langle H_3 \rangle $) are identical except a small
part which is different. The reasons for this difference are
analyzed in the above paragraph. While the latitudinal structure and
variation with the solar cycle of $ \langle H_1 \rangle $ and $
\langle H_2 \rangle $ ( $ \langle H_3 \rangle $ and $ \langle H_6
\rangle $) are much different which confirms again the absence of
isotropy for the observable current helicity in solar active
regions. We can also see that in some periods and latitudinal
intervals the helicity parts satisfy the hemispheric sign rule
(negative/positive sign in the North/South hemispheres) well, e.g.,
1992--1994 and 2001--2003 for $ \langle H_1 \rangle $ in southern
hemisphere. The fractions of magnetograms following the hemispheric
sign rule are 46.2\% for $ \langle H_1 \rangle $, 61.9\% for $
\langle H_2 \rangle $, 60.2\% for $ \langle H_3 \rangle $ and 47.2\%
for $ \langle H_6 \rangle $ in northern hemisphere respectively,
while in southern hemisphere they are 57.9\% for $ \langle H_1
\rangle $, 55.4\% for $ \langle H_2 \rangle $, 54.4\% for $ \langle
H_3 \rangle $ and 57.1\% for $ \langle H_6 \rangle $ respectively.

\begin{figure} 
\centering \resizebox{0.48\textwidth}{!}
{\includegraphics[]{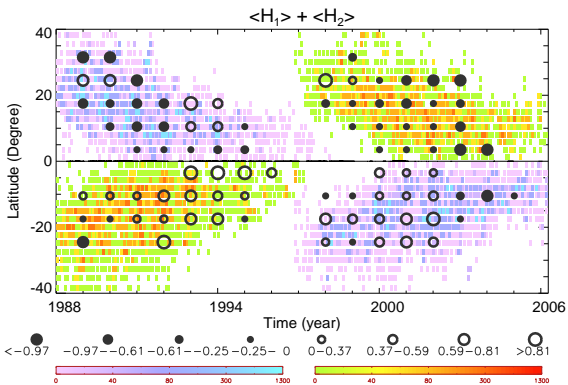}}
\caption{%
Evolution of the observationally available helicity defined as a sum
of two parts $ \langle  H_1 \rangle +\langle H_2 \rangle $  with the
solar cycle. The unit of helicity is $10^{-3}G^{2}/m$.} \label{fig8}
\end{figure}

The sum $ \langle H_{cz} \rangle = \langle H_1 \rangle + \langle H_2
\rangle $ usually used in the past study in Figure \ref{fig8} shows
some anti-symmetry with respect to the solar equator and the
hemispheric sign rule is well pronounced, this figure is almost the
same as Figure~2 in \citet{Zhang10}. The use of the sum of the other
two helicity parts $H_6$ and $H_3$, respectively, would produce
visually identical result, and not shown here. The fractions of
magnetograms following the hemispheric sign rule for $ \langle
H_{cz} \rangle$ are 58.9\% in northern hemisphere and 61.7\% in
southern hemisphere respectively. We can see some regular inversions
of hemispheric sign rule within isolate ranges of latitudes near the
beginning and the end of the solar cycle for $ \langle H_{cz}
\rangle$. We note that during these phases the signs of pair of
helicity parts ($ \langle H_1 \rangle $ and $ \langle H_2 \rangle $)
are often the same and violate the hemispheric sign rule, which
makes their joint contribution to the reversal of hemispheric sign
rule.

\begin{table*}
\begin{center}
\caption{Summary of the relative degree of anisotropy. The unit of
$H_{1}$ and $H_{2}$ is $10^{14} G^{2}m$. } \label{tab:xu2}
\begin{tabular}{ccccccccccc}
\hline
Instr. &NOAA & date & Time & coordinate & $  H_1$ & $ H_2$ &  $q_{12}$ & $q_{12}^a$  \\
\hline
SMFT &8898 & 2000.03.08& 01:37 &S13.0W7.0 & $-0.0493$ & 0.0310  & 0.8348& 0.8325 \\
SMFT &6659 & 1991.06.09 & 05:29 & N28.6E4.5 & $-1.4978$ & $-2.5897$ & 0.7823& 0.7794 \\
HMI  & 11158 & 2011.02.14 & 23:47 & S20W17 & 0.0793 & 0.7739 & 0.7472 & 0.7472 \\
\hline
\end{tabular}
\end{center}
\end{table*}

\section{Discussion}
     \label{S-Discussion}

In order to establish the degree of anisotropy quantitatively, we
take 3 active regions which used for calculating boundary integral
in section 3 for examples. We estimate the relative degree of
anisotropy, for example, as a relative anisotropy imbalance between
the two parts of helicity using some norm as following:

\begin{equation}
q_{12}={\|h_1-h_2\| \over \|h_1\|+\|h_2\|},
\end{equation}
where the norm $\|h\|$ can be computed as
\begin{equation}
\|h\|= \sqrt{ {\int(w|h|^2dxdy) \over \int(wdxdy)}},
\end{equation}
where $w$ is weight factor as follow: when we integrate in the
magnetogram using all pixels, then $w=1$. Alternatively, we can use
only pixels where the signal is greater than cut-off noise levels
($|B_{z}| > 20 G$ and $B_{t} > 100 G$) as we usually do so in
helicity statistical studies ($w=1$ if greater, or $w=0$ if less
than noise level). The value of $q_{12}$ is between 0 to 1. The
results are listed in Table~\ref{tab:xu2}. $q_{12}$ is calculated
using all pixels in the magnetogram, $q_{12}^a$ is only using the
pixels where the signal is greater than cut-off noise levels. From
this table, the order of $q_{12}$ is from 0.74 to 0.83 and it is not
affected much by the noise, which is close to 1, it says that the
observed quantities are extremely anisotropic.

We may further speculate about the sources of anisotropy of current
helicity.  One effect which sounds trivial in the solar photosphere
would be vertical stratification of convection. We, however, are
inclined to discuss this issue as the parts of helicity involving
vertical derivatives of the magnetic field are not observationally
available. The difference in properties and behaviour of $ H_1$ and
$ H_2$ is likely a manifestation of the effect of rotation as one
contains derivatives with respect to azimuthal (collinear to
rotation), and the other meridional direction (perpendicular to
rotation). The other source of anisotropy could be the effect of
large-scale magnetic field. Two components (azimuthal and radial) of
the solar cyclic magnetic field are anti-symmetric over the equator
and the other meridional component is symmetric over the solar
equator.

The complete use of the assumption of local statistical isotropy
means also the equality between these two groups of the three parts
of helicity in  equation~\ref{h146} and~\ref{h235}. Two of them, $
H_4 $ and $ H_5 $, however, can not be calculated from the
observational vector magnetograph data at the solar photosphere.
Therefore, we can not check the condition of local isotropy of solar
turbulence in active region by observationally available data
completely. On the Other hand, there are several factors which
affect the precision of the calculated helicity parts, such as that
the accuracy of transversal field is lower than the one of
longitudinal field, the magneto-optical effects, calibration of
magnetic fields etc. We cannot completely estimate the influence of
these factors on the helicity parts although we used some data
reduction method to reduce it. This topic requires further
investigation.

\section{Conclusion}
     \label{S-Conclusion}

We have studied the distribution and properties of the parts of current helicity from observation and simulation. The main conclusions are as following:

(1) The simulation results show that the means of the six helicity
parts statistically coincide with each other for the isotropic case,
while for the anisotropic helical case it is different.

(2) The distribution of the observed helicity parts is similar with
the helical anisotropic case in simulations. This shows that the
distribution of the observed helicity parts is anisotropic over
active region scales. Assumptions of local homogeneity and isotropy
in computation of observational proxies of helicity require further
analysis in the light of our findings.

(3) The four observable helicity parts are equal in pairs owing to
the magnetic field is weak at the boundaries of magnetogram, but
there is a large difference between different pairs which may be
caused by the anisotropy of the current helicity density in active
region. Both the pairs of the helicity parts follow the hemispheric
sign rule with certain exceptions in periods and hemispheres but the
sums of the two parts follow the hemispheric sign rule more robust
and uniform than the individual parts of helicity alone.

More theoretical modeling of anisotropy in solar-like turbulence is
required to understand these statistical results. Our simple
simulations have shown that the anisotropy may be present in several
directions and not only in the direction of vertical stratification.
This is confirmed by several examples of observational data as well
as their statistical analysis. Other sources of anisotropy such as
inhomogeneity or rotation, and furthermore presence of the
large-scale {\it mean} magnetic field which alternates in sign with
every 11-year sunspot cycle may be possible explanations of these
results.

\section*{Acknowledgments}
This work is supported by the National Natural Science Foundation of
China (Grant Nos. U13311044,1174153, 11173033, 11178005, 11221063,
11203036, 11373040, 11303052, 11303048, 11125314, 11473039),
Knowledge Innovation Program of The Chinese Academy of Sciences
(Grant No. KJCX2-EW-T07), National Basic Research Program of China
(Grant No. 2011CB811401), Grant No. XDA04060804-02, and the Young
Researcher Grant of National Astronomical Observatories, Chinese
Academy of Sciences. This work is a result of long term cooperation
between the Chinese and Russian teams supported by NSFC of China and
RFBR of Russia joint grant (NSFC number 1141101089 and RFBR number
15-52-53125). K.K. would like to acknowledge Chinese Academy of
Sciences Visiting Professorship grant.



\begin{thebibliography}{}
\bibitem[\protect\citeauthoryear{Abramenko et al.}{1996}]{Abramenko96}
Abramenko, V.I., Wang, T.J., Yurchishin, V.B., 1996, {\solphys}
\textbf{168}, 75.

\bibitem[\protect\citeauthoryear{Ai \& Hu}{1986}]{Ai86}
Ai, G.X., Hu, Y.F., 1986, {\ Acta Astron. Sinica} \textbf{27}, 173.

\bibitem[\protect\citeauthoryear{Bao \& Zhang}{1998}]{Bao98}
Bao, S.D., Zhang, H.Q., 1998, {\apjl} \textbf{496}, L43.

\bibitem[\protect\citeauthoryear{Babcock}{1961}]{Babcock1961}
Babcock, H., 1961, {\apj}, 133, 572.

\bibitem[\protect\citeauthoryear{{Berger}}{2003}]{Berger03b}
Berger M. A., 2003, in Ferriz-Mas, A., N{\'u}{\~n}ez, M. (eds.),
    \textit{Advances in Nonlinear Dynamics}, Taylor and Francis Group,
    London, 345.

\bibitem[\protect\citeauthoryear{{Berger} and {Field}}{1984}]{BergerF84b}
Berger M. A., Field G. B., 1984, \textit{J. Fluid. Mech.}
\textbf{147}, 133.

\bibitem[\protect\citeauthoryear{Brown et al.}{1999}]{Brown99b} Brown M., Canfield R., Pevtsov A., 1999,
Magnetic Helicity in Space and Laboratory Plasmas, Geophys. Mon.
      Ser. 111, AGU.

\bibitem[\protect\citeauthoryear{Christensson et al.}{2001}]{Christensson2001}
{Christensson}, M. and {Hindmarsh}, M. and {Brandenburg}, A.
{\pre} \textbf{64}, 056405 (2001).

\bibitem[\protect\citeauthoryear{Dupont et al.}{2007}]{Dupont07b}
Dupont, J.-C., Schmidt, F., Koutny, P., 2007, \solphys{}
\textbf{323}, 965.

\bibitem[\protect\citeauthoryear{Gao et al.}{2008}]{Gao08}
Gao, Y., Su, J., Xu, H. and Zhang, H., 2008, {\it Mon. Not. R.
Astron. Soc.}, 386, 1959.

\bibitem[\protect\citeauthoryear{Goldreich and Sridhar}{1997}]{Goldreich97}
Goldreich P., Sridhar S., 1997, {\apj} \textbf{485}, 680.

\bibitem[\protect\citeauthoryear{Hagino and Sakurai}{2004}]{Hagino04}
Hagino, M., Sakurai, T.: 2004, {\pasj} \textbf{56}, 831.

\bibitem[\protect\citeauthoryear{Iroshnikov}{1963}]{Iro63}
Iroshnikov, P. S. 1963, AZh, 40, 742

\bibitem[Kolmogorov(1941)]{Ko41}
Kolmogorov, A.N., 1941, Dokl. A N SSSR, 30, 299

\bibitem[\protect\citeauthoryear{Kraichnan}{1965}]{Krai65}
Kraichnan, R. H. 1965, Phys. Fluids, 8, 1385

\bibitem[\protect\citeauthoryear{Leighton}{1969}]{Leighton69}
Leighton, R., 1969, {\apj}, 156, 1.

\bibitem[\protect\citeauthoryear{Parker}{1955}]{Park55}
Parker, E., {\apj} \textbf{122}, 293.

\bibitem[\protect\citeauthoryear{Pevtsov}{1994}]{pev94}
Pevtsov, A.A., Canfield, R.C., Metcalf, T.R., 1994, {\apj},
425, L117.

\bibitem[\protect\citeauthoryear{Seehafer}{1990}]{Seehafer90}
Seehafer, N. 1990, {\solphys} \textbf{125}, 219.

\bibitem[\protect\citeauthoryear{Wang, Xu, and Zhang}{1994}]{Wang94}
Wang T.J., Xu A.A., Zhang H.Q. 1994, {\solphys}, 155, 99.

\bibitem[\protect\citeauthoryear{Zhang et al.}{2010}]{Zhang10}
Zhang Hongqi, Sakurai T., Pevtsov A, Gao Yu , Xu Haiqing, Sokoloff
D. D. and Kuzanyan K., 2010, {\it Mon. Not. R. Astron. Soc.}
\textbf{402}, L30.

\bibitem[\protect\citeauthoryear{Zhang et al.}{2012}]{Zhang12}
Zhang H., Moss D., Kleeorin N., Kuzanyan K., Rogachevskii I.,
Sokoloff D., Gao Y., and Xu H., 2012, {\apj} \textbf{751}, 47.



\end{thebibliography}
\end{document}